\documentclass[a4paper,12pt]{article}
\usepackage{amssymb,amsmath,amsfonts,amsthm}
\usepackage{verbatim,graphicx}
\usepackage{pstool}

\usepackage{pgfplots}
\pgfplotsset{compat=newest}
\usetikzlibrary{plotmarks}
\usetikzlibrary{external}
\usepackage{grffile}
\usepackage{apacite}

\newtheorem{thm}{Theorem}
\newtheorem{cor}[thm]{Corollary}

\newcommand{\bs}{\boldsymbol}

\newcommand{\var}{\mathrm{var}}
\newcommand{\re}{\mathrm{e}}
\renewcommand{\epsilon}{\varepsilon}
\newcommand{\pmaj}{p_{\mathrm{maj}}}
\newcommand{\rhoUK}{\rho_{\mathrm{UK}}}
\newcommand{\rhoPak}{\rho_{\mathrm{Pak}}}

\newcommand{\hfigwidth}{6.5cm}

\numberwithin{equation}{section}

\begin{document}
\title{Evaluation of vaccination strategies for SIR epidemics on random networks incorporating household structure}
\author{Frank Ball\thanks{School of Mathematical Sciences, University of Nottingham, UK.} \and
David Sirl\footnotemark[1]}
\date{October 2016}

\renewcommand{\BCBT}{}
\renewcommand{\BCBL}{}

\maketitle

\begin{abstract}
This paper is concerned with the analysis of vaccination strategies in a stochastic SIR (susceptible $\to$ infected $\to$ removed) model for the spread of an epidemic amongst a population of individuals with a random network of social contacts that is also partitioned into households.
Under various vaccine action models, we consider both household-based vaccination schemes, in which the way in which individuals are chosen for vaccination depends on the size of the households in which they reside, and acquaintance vaccination, which targets individuals of high degree in the social network.
For both types of vaccination scheme, assuming a large population with few initial infectives, we derive a threshold parameter which determines whether or not a large outbreak can occur and also the probability and fraction of the population infected by such an outbreak.
The performance of these schemes is studied numerically, focusing on the influence of the household size distribution and the degree distribution of the social network.
We find that acquaintance vaccination can significantly outperform the best household-based scheme if the degree distribution of the social network is heavy-tailed.
For household-based schemes, when the vaccine coverage is insufficient to prevent a major outbreak and the vaccine is imperfect, we find situations in which both the probability and size of a major outbreak under the scheme which minimises the threshold parameter are \emph{larger} than in the scheme which maximises the threshold parameter.
\end{abstract}


\section{Introduction and description of results}
\label{secintro}

Mathematical models for the spread of infectious disease have much to offer in terms of understanding past outbreaks, predicting likely behaviours of future outbreaks and predicting the effect of interventions or mitigating strategies. In the last decade or two there has been considerable interest and work on network epidemic models. These involve supplanting the traditional assumption of homogeneous mixing of homogeneous individuals with some random graph structure, with specific interest in being able to control the degree distribution, reflecting the varying numbers of people with which different individuals tend to interact. Other structures, for example including households and stratification of populations, have been studied for longer~\cite{Bartoszynski1972,BalMolSca1997,Watson1972,Scalia-Tomba1986}; but structures with a `social network' type of interpretation start around the turn of the millenium with the works of~\citeA{DieDeJMet1998,Andersson1997,Andersson1998,Newman2002}. In this and most other papers in the field we typically have in mind an infection spreading through a human population. However, much the same ideas apply to mathematical models of a variety of other motivating applications, such as the spread of rumours or information through human populations, infection or information spread through a population of other animals and virus spread through a network of computers.

In this paper we build on the model of~\citeA{BalSirTra2009,BalSirTra2010} which includes household and network structure to include vaccination, with some emphasis on so-called acquaintance vaccination \cite{CohHavben2003,BriJanMar2007} as elucidated in~\citeA{BalSir2013} in a model without household structure. In the model of \citeA{BalSirTra2010}, a population of fixed size is given social network structure via the configuration model random graph (see e.g.\ \citeA{Bollobas1980,NewStrWat2001,Newman2002}) and the population is also partitioned into households (see e.g.\ \citeA{BalMolSca1997}). A stochastic SIR (Susceptible--Infective--Removed) epidemic model is then defined on this population structure. 
A first quantity of interest in this model is the final size, which is the (random) number of initial susceptibles that are infected at some point during the epidemic. In line with much of modern stochastic epidemic theory, one can use branching process approximations to prove a threshold theorem (valid in the large population limit) which determines whether the infection will necessarily die out relatively quickly, resulting in a small final size, or whether it is possible for the epidemic to take off and infect a substantial fraction of the population. These methods also yield approximations for the probability that a supercritical epidemic will take off and using closely related methods one can also study final size properties of such a large outbreak.

%, in which we begin with a completely susceptible population except for a single initial infective. Infective individuals infect susceptible neighbours in the network (individuals in the same household and neighbours in the random graph) during a random infectious period and then become removed and play no further part in the epidemic. This continues until there are no infective individuals remaining, at which point some of the population will have been infected and ultimately be removed and some of the initial susceptibles will still be susceptible.

This paper provides tools for studying the effect of introducing vaccination (or some other action which is implemented in advance of the spread of the epidemic) into this model. %By vaccination we mean any pre-emptive action which affects the susceptibility or infectivity of an individual in the population. Crucially, the intervention must be before the startof the epidemic process, we do not allow for on-line responses such as contact tracing.
\emph{Households-based} vaccination schemes are those that can be described in terms of the distribution of the number of vaccinated individuals in households of size $n$, for every household size $n$ in the population. This includes as special cases the situation when we vaccinate individuals who are chosen uniformly at random from the population (the distributions are binomial) and vaccinating households at random (the distributions are concentrated at 0 and $n$). Optimal schemes in this context often resemble the equalising strategy where one vaccinates preferentially in larger households, there being more of a herd immunity effect available to exploit in those larger households on top of the direct protection of vaccinated individuals. \emph{Acquaintance vaccination} schemes exploit the heterogeneity of individuals' connectivities in a network to preferentially target better-connected individuals for vaccination. Instead of vaccinating individuals chosen from the population in some way, one samples individuals and then vaccinates their friends---their acquaintances in the network. This exploits the so-called friendship paradox, the observation that, for most people, their friends on average have more friends than they do \cite{Feld1991}.

Our main theoretical results are the calculation of the asymptotic final size quantities (i.e. characterization of appropriate branching process approximations) when we include vaccination in the \citeA{BalSirTra2010} household-network model. This extends the results of~\citeA{BecSta1997} and \citeA{BalLyn2002,BalLyn2006} on the standard households model to have network-based (rather than homogeneous mixing) casual contacts. It also extends results of~\citeA{BalSir2013} on acquaintance vaccination in a population with network (but not household) structure. We also explore the model numerically and find that there can be substantial differences in the performance of the different vaccine allocation strategies.

%(We typically think of infections being more likely to occur between individuals in the same household than between individuals connected in the random graph.)

The remainder of the paper is structured as follows. In Section~\ref{secmodthrvac} we specify our models for the population structure and evolution of the epidemic, then outline the analysis of the final outcome of the epidemic and lastly introduce the model we use for the action of a vaccine on individuals who receive it. In Section~\ref{sechousevac} we consider the effect of households-based vaccination, including optimal households-based strategies, and in Section~\ref{secacq} we consider the effect of acquaintance vaccination, analysing the same final outcome properties of the epidemic. Some exploration of the behaviour of the model (mainly numerical) is presented in Section~\ref{secbehaviour}. Lastly we offer some concluding remarks in Section~\ref{secconclusion}. Details of several of the longer calculations from Sections~\ref{secmodthrvac}--\ref{secacq}, as well as some standard results for small homogenously mixing populations (households in our context), are given in the appendix.

\section{Model, threshold behaviour and vaccination}
\label{secmodthrvac}
\subsection{Model}
\label{subsecmod}
The model under consideration in this paper is that of \citeA{BalSirTra2010} for the spread of an SIR epidemic on a finite random network incorporating household structure.  We assume that the population consists of $N$ individuals and is partitioned into $m$ households, of which $m_n$ are of size $n$ ($n=1,2,\ldots$).  Thus $m=\sum_{n=1}^{\infty} m_n$ and $N=\sum_{n=1}^{\infty} nm_n$.  The network of possible global (i.e.\ between-household) contacts is constructed using the configuration model (with random, rather than specified, degree sequence).  Thus each individual is assigned a number of `half-edges' independently, according to an arbitrary but specified discrete random variable $D$ having mass function $P(D=k)=p_k$ ($k=0,1,\ldots$), and then all of the half-edges are paired up uniformly at random to form the edges in the graph describing the global network.  If the total number of half-edges is odd, we ignore the single leftover half-edge.

Our analysis is asymptotic as the number of households $m \to \infty$.  We require that, as $m \to \infty$, $m_n / m \to \rho_n$ ($n=1,2,\ldots$), where $(\rho_1, \rho_2 , \ldots$) is a proper probability distribution (i.e.\ $\sum_{n=1}^{\infty} \rho_n = 1$) having finite mean $\mu_H = \sum_{n=1}^{\infty} n \rho_n$.  Thus $\mu_H$ is the mean household size in the limiting population.  We also require that $\mu_D = E[D]$ is finite.  These assumptions are sufficient for our analysis.  If we make the stronger assumptions that $\sigma_D^2 = \var (D)$ and $\sum_{n=1}^{\infty} n^2 \rho_n$ are both finite, then parallel edges and self-loops, between either individuals or households, become sparse in the global network as $n \to \infty$.

The epidemic is initiated by a single individual, chosen uniformly at random from the population, becoming infected, with the other individuals in the population all assumed to be susceptible.  The infectious periods of different infectives are each distributed according to a random variable $I$, having an arbitrary but specified distribution.  Throughout its infectious period, a given infective makes infectious contact with any given member of its household at the points of a Poisson process having rate $\lambda_L$ and with any given global neighbour at the points of a Poisson process with rate $\lambda_G$.  (Note that $\lambda_L$ and $\lambda_G$ are both individual to individual contact rates.)  If an individual so contacted is susceptible then it becomes infected, otherwise the contact has no effect.  Contacted susceptibles are immediately able to infect other individuals, i.e.\ there is no latent period.  An infective individual becomes removed at the end of its infectious period and plays no further role in the epidemic.  All infectious period, global degrees and Poisson processes are assumed to be mutually independent.  The epidemic ceases as soon as there is no infective in the population.

For ease of exposition we have assumed that there is no latent period and that the epidemic is started by a single infective chosen uniformly at random from the population.  As explained in \citeA{BalSirTra2010}, these assumptions may be relaxed without compromising mathematical tractability.  In particular, our results are related to the final outcome of the epidemic, the distribution of which is invariant to very general assumptions concerning a latent period (see e.g.\ \citeA{PelFerFra2008}).

\subsection{Threshold behaviour}
\label{subsecthreshold}

\subsubsection{Early stages of epidemic}
\label{subsecearlystages}

Recall that the global network is formed by pairing up the half-edges uniformly at random.  It follows that in the early stages of an epidemic the probability that a global contact is with an individual residing in a previously infected household is small, indeed it is zero in the limit as $m \to \infty$.  Thus, in the early stages of an epidemic, the process of infected households can be approximated by a branching process.  The individuals in this branching process correspond to infected households in the epidemic process, and the offspring of a given individual in the branching process are all households that are contacted globally by members of the local (within-household) epidemic in the parent household.

The offspring distribution for this branching process is usually different in the initial generation than in all subsequent generations.  The number of global neighbours that the initial infective in the initial infected household may infect is distributed according to $D$, as that individual is chosen uniformly at random from the entire population.  The number of global neighbours that the initial infective in any subsequent infected household (in the branching process approximation) may infect is distributed according to $\tilde{D}-1$, where $\tilde{D}$ is the degree of a typical neighbour of a typical individual in the network.  The $-1$ arises because such an initial infective has been infected through the global network, so one of its neighbours (i.e.\ its infector) is not available for further infection.  Note that a given half-edge is $k$ times as likely to be paired with a half-edge emanating from an individual with degree $k$ than with one emanating from an individual with degree 1, so $P(\tilde{D}=k)=\mu_D^{-1} k p_k$ ($k=1,2,\ldots$).  The distributions of $D$ and $\tilde{D}-1$ are equal if and only if $D$ has a Poisson distribution. For any non-negative integer valued random variable $X$, we denote its probability generating function (PGF) by $f_X$, so $f_X(s) = E[s^X]$ ($0 \leq s \leq 1$). We note for future reference that $f_{\tilde{D}-1}(s) = f'_D(s)/\mu_D$.

%\footnote{Added new sentence. Haven't otherwise mentioned this relationship.}Using the notation $f_X$ for the probability generating function (PGF) of a non-negative integer valued random variable $X$, we note for future reference that $f_{\tilde{D}-1}(s) = f'_D(s)/\mu_D$.

The approximation of the early stages of the epidemic process by the above branching process is made mathematically fully rigorous in \citeA{BalSirTra2009} and \citeA{BalSir2012}.  The latter shows that a sequence of epidemic processes, indexed by $m$, and the approximating branching process can be constructed on the same probability space so that, as $m \to \infty$, the number of households ultimately infected in the epidemic process converges almost surely to the total progeny of the branching process.  Thus, provided $m$ is large, whether or not the epidemic can become established and lead to a major outbreak is determined by whether or not the branching process is supercritical.

Let $C$ and $\tilde{C}$ be random variables describing the number of offspring of the initial and a typical subsequent individual, respectively, in the branching process.  Then standard branching process theory (e.g.\ \citeNP{HacJagVat2005}, Theorem~5.2), gives that the extinction probability of the branching process is strictly less than one if and only if $R_{\ast} = E[\tilde{C}] > 1$.  Thus $R_{\ast}$ serves as a threshold parameter for the epidemic model.  We now outline the calculation of $R_{\ast}$.  Further details are given in \cite{BalSirTra2010}.

First note that, since the degree and household size of an individual are independent, the probability that a typical globally infected individual resides in a household of size $n$ is given by $\tilde{\rho}_n = \mu_H^{-1} n \rho_n$ ($n=1,2,\ldots$).  (An individual chosen uniformly at random from the population is $n$ times as likely to reside in a given household of size $n$ than in a given household of size 1.)  Thus,
\begin{equation}
E[\tilde{C}]=\sum_{n=1}^{\infty} \tilde{\rho}_n E[\tilde{C}^{(n)}],\label{ECtilde}
\end{equation}
where $\tilde{C}^{(m)}$ is the number of global infections emanating from a typical size-$n$ single-household epidemic initiated by a single infective who is infected through the global network.  Consider such a size-$n$ single-household epidemic.  Label the household members $0,1,\ldots,n-1$, where $0$ is the initial infective, and write
\begin{equation}
\tilde{C}^{(n)}=C_0 + \sum_{i=1}^{n-1} \chi_i C_i ,\label{Ctilden}
\end{equation}
where $\chi_i=1$ if individual $i$ is infected by the single-household epidemic, otherwise $\chi_i=0$, and $C_i$ is the number of global infections made by individual $i$ (assuming it is infected).  Let $T^{(n)}=\sum_{i=1}^{n-1} \chi_i$ be the final size of the single-household epidemic, not including the initial infective, and $\mu^{(n)} (\lambda_L)=E[T^{(n)}]$.  (A formula for $\mu^{(n)} (\lambda_L)$ is given in equation (\ref{munlambdaL}) in Appendix~\ref{AppMeanLocalFS}.) Whether or not a given individual, $i$ say, is infected by the single-household epidemic is independent of its infectious period, so $\chi_i$ and $C_i$ are independent.  Thus, taking expectations of (\ref{Ctilden}) and exploiting symmetries yields
\begin{equation}
E[\tilde{C}^{(n)}] = E[C_0] + \mu^{(n)} (\lambda_L ) E[C_1].\label{ECtilden}
\end{equation}

The probability that individual $0$ infects a given global neighbour is $p_G = 1 - E[ \re^{-\lambda_G I}]$ and the number of uninfected global neighbours of individual $0$ is distributed according to $\tilde{D}-1$.  Thus $E[C_0]=p_G \mu_{\tilde{D}-1}$, where $\mu_{\tilde{D}-1}=E[\tilde{D}-1]=\mu_D + \frac{\sigma_D^2}{\mu_D} - 1$.  Similarly, $E[C_1]=p_G \mu_D$, since the number of uninfected global neighbours of individual $1$ is distributed according to $D$.  Substituting these results into (\ref{ECtilden}), and then into (\ref{ECtilde}), yields
\begin{equation}
R_{\ast} = p_G \left[ \mu_{\tilde{D}-1} + \mu_D \sum_{n=1}^{\infty} \tilde{\rho}_n \mu^{(n)} (\lambda_L) \right] .\label{Rstar}
\end{equation}

Let $\pmaj$ be the probability that a major outbreak occurs.  Then standard branching process theory (e.g.\ \citeA[Theorem 5.2]{HacJagVat2005}), shows that $\pmaj=1-f_C (\sigma)$, where $\sigma$ is the smallest solution of $f_{\tilde{C}} (s)=s$ in $[0,1]$.  Note that, analagous to~(\ref{ECtilde}), $f_C(s)=\sum_{i=1}^{\infty} \tilde{\rho}_n f_{C^{(n)}}(s)$ and $f_{\tilde{C}} (s) = \sum_{n=1}^{\infty} \tilde{\rho}_n f_{\tilde{C}^{(n)}} (s)$. Details of the calculation of $f_{C^{(n)}}$ and $f_{\tilde{C}^{(n)}}$ are given in Appendices~\ref{AppKeyResult} and~\ref{AppNoVac}. As noted in \citeA[Section 3.2]{BalSirTra2010}, these calculations are much simpler in the case where the infectious period is constant.

\subsubsection{Final outcome of major outbreak}
\label{subsecfinaloutcome}
We now consider the fraction of the population that are ultimately infected by an epidemic that takes off.  The key tool we use is the \emph{susceptibility set} \cite{Ball2000ResRep,BalLyn2001,BalNea2002}, which we now describe.  Let $\mathcal{N}=\{1,2,\ldots,N\}$ denote the entire population of $N$ individuals.  For each $i \in \mathcal{N}$, by sampling from the infectious period distribution and then the relevant Poisson processes for local and global contacts, draw up a (random) list of individuals $i$ would make infectious contact with if it was to become infected.  Then construct a random directed graph on $\mathcal{N}$, in which for any pair, $(i,j)$ say, of individuals there is an arc from $i$ to $j$ if and only if $j$ is in $i$'s list.  The susceptibility set of a given individual, $i$ say, consists of those individuals from which there is a chain of arcs to $i$ in the graph (including $i$ itself).  Note that any given individual is ultimately infected by the epidemic if and only the initial infective belongs to its susceptibility set.

As for the early stages of an epidemic, we can approximate the susceptibility set of an individual, $i$ say, chosen uniformly at random from the population, by a households-based branching process.  We first consider $i$'s \emph{local} susceptibility set, i.e.\ the susceptibility set obtained when only local (within-household) contacts are considered.  Suppose that this local susceptibility set has size $M^{(n)} +1$, where $n$ denotes the size of $i$'s household.  (The probability mass function of $M^{(n)}$ is given by equation (\ref{susmass1}) in Appendix~\ref{AppSusset}.) Let $B$ be the number of individuals, who are not in $i$'s household, that in the random directed graph have an edge leading directly \emph{to} one of the $M^{(n)} +1$ individuals in $i$'s local susceptibility set.  Each individual in $i$'s local susceptibility set has global degree distributed independently according to $D$, and, as $m \to \infty$, each global neighbour of $i$'s local susceptibility set enters $i$'s susceptibility set independently with probability $p_G$.  Thus, taking expectations with respect to $i$'s household size,
\begin{equation}
f_B (s) = \sum_{n=1}^{\infty} \tilde{\rho}_n f_{B^{(n)}} (s), \label{pgfB}
\end{equation}
where
\begin{equation}
f_{B^{(n)}} (s) = f_D (1-p_G + p_G s) f_{M^{(n)}} (f_D (1-p_G + p_G s)).\label{pgfBn}
\end{equation}

The above $B$ individuals form the first generation of our approximating branching process.  We next consider each of these $B$ individuals in turn, construct the local susceptibility sets in their respective households (which are distinct with probability tending to one as $m \to \infty$) and then examine the global neighbours of these local susceptibility sets to obtain the second generation of the approximating branching process, and so on.  Note that the initial individual in each of these $B$ local susceptibility sets has degree distributed according to $\tilde{D}$, so, as previously, the offspring distribution for the branching process is different for the initial individual than for all subsequent individuals.  If we let $\tilde{B}$ denote the offspring random variable of a typical non-initial individual, then arguing as in the derivation of (\ref{pgfB}),
\begin{equation}
f_{\tilde{B}} (s) = \sum_{n=1}^{\infty} \tilde{\rho}_n f_{\tilde{B}^{(n)}} (s), \label{pgfBtilde}
\end{equation}
where
\begin{equation}
f_{\tilde{B}^{(n)}} (s) = f_{\tilde{D}-1} (1-p_G + p_G s) f_{M^{(n)}} (f_D (1-p_G + p_G s)).\label{pgfBtilden}
\end{equation}

The probability that the approximating branching process survives (i.e.\ does not go extinct) is given by $z=1-f_B (\xi)$, where $\xi$ is the smallest solution of $f_{\tilde{B}} (s)=s$ in $[0,1]$.  It is straightforward to show that $E[\tilde{B}]=E[\tilde{C}]$, so $z > 0$ if and only if $R_{\ast}>1$. Moreover, if $R_{\ast} > 1$ then $z$ is the expected proportion of the population that is ultimately infected by a major outbreak in the limit as $m \to \infty$; see \citeA{BalSirTra2009} for a formal proof when all the households have the same size.  Furthermore, the method of proof of \citeA[Theorem~3.4]{BalSirTra2014} can be adapted to show that, as $m \to \infty$, the proportion of the population that is ultimately infected by a major outbreak converges in probability to $z$.  Thus we refer to $z$ as the relative final size of a major outbreak.

\subsection{Vaccination}
\label{subsecvacc}
In modelling vaccination there are two distinct aspects that must be modelled: (i) who gets vaccinated and (ii) what happens to those who are vaccinated. Vaccine \emph{allocation} models (addressing the former issue) are our focus in this paper. We now outline the vaccine \emph{action} models (addressing the latter aspect) that we allow for in our analysis.

We use a model for vaccine action, proposed by \citeA{BecSta1998}, in which the vaccine response of an individual who is vaccinated is described by a random vector $(A,B)$, where $A$ denotes the relative susceptibility (compared to an unvaccinated individual) and $B$ the relative infectivity if the vaccinated individual becomes infected.  Thus all Poisson processes concerning potential infection of the individual have their rates multiplied by $A$ and the Poisson processes governing the contacts the individual makes, if infected, have their rates multiplied by $B$.  The vaccine responses of distinct vaccinees are assumed to be mutually independent.  Within this framework we consider two special cases, the \emph{all-or-nothing} and the \emph{non-random} vaccine responses.

The all-or-nothing model (see e.g.\ \citeA{HalHabLon1992}) is obtained by setting $P((A,B)=(0,0))=1-P((A,B)=(1,1))=\epsilon$, so vaccinated individuals are rendered completely immune with probability $\epsilon$, otherwise the vaccine has no effect.  The non-random model (see e.g.\ \citeA{BalLyn2006}) assumes that $P((A,B)=(a,b))=1$, for some $(a,b)$, so all vaccinated individuals respond identically.  An important special case is the \emph{leaky} model (see e.g.\ \citeA{HalHabLon1992}), when $b=1$, so vaccination does not affect an individual's ability to transmit the disease if they become infected. Setting $\epsilon=1$ in the all-or-nothing model or $a=b=1$ in the non-random model yields a \emph{perfect} vaccine response; that is one in which all vaccinated individuals are rendered completely immune.

%%%%%%

\section{Households based vaccination} 
\label{sechousevac}
\subsection{Introduction} 
\label{subsechousevacintro}
In this section we consider vaccine allocation strategies based on household size and determine their impact on $R_{\ast}$, $\pmaj$ and $z$ under the all-or-nothing and non-random vaccine action models.  For $n=1,2,\ldots$ and $v=0,1,\ldots,n$, let $x_{nv}$ denote the proportion of households of size $n$ that have $v$ members vaccinated.  Let $p_V$ denote the proportion of the population that is vaccinated, i.e.\ the \emph{vaccination coverage}.  Then $p_V$ is also the probability that an individual chosen uniformly at random from the population is vaccinated.  Conditioning on the size of such an individual's household yields
\begin{equation}
p_V = \sum_{n=1}^{\infty} \tilde{\rho}_n \sum_{v=0}^n \frac{v}{n} x_{nv}.\label{pv}
\end{equation}
We derive results for an arbitrary but specified vaccine allocation however in the numerical studies we consider four allocation schemes: uniformly chosen households, uniformly chosen individuals, `best' and `worst'.  In the uniformly chosen households scheme, households are chosen uniformly at random and all of their members are vaccinated.  Thus, if the vaccination coverage is $p_V$, $x_{nv}=p_V \delta _{vn}+(1-p_V)\delta_{v0}$ ($n=1,2,\ldots$; $v=0,1,\ldots ,n$), where $\delta_{vk}=1$ if $v=k$ and $\delta_{vk}=0$ if $v \neq k$.  In the uniformly chosen individuals scheme, individuals are chosen uniformly at random and vaccinated, so
$x_{nv} = \binom{n}{v} p_V^v (1-p_V)^{n-v}$ ($n=1,2,\ldots$; $v=0,1,\ldots,n$).  The best and worst schemes are the allocations that make $R_{\ast}$ respectively as small as possible and as large as possible, for a given vaccination coverage.

\subsection{All-or-nothing vaccine action} %3.2
\label{subsechousevacaon}
To analyse the consequences of a vaccination scheme using an all-or-nothing vaccine action model it is convenient to use the concept of a potential infectious global contact.  Consider a given infected individual and a given global neighbour of that infected individual.  Then that neighbour is a potential infectious global contact of the infective, if it is in that infective's list of individuals it makes infectious contact with (see the start of Section~\ref{subsecfinaloutcome}).  The potential infectious global contact becomes an actual infectious global contact if either the neighbour is unvaccinated or it is vaccinated but the vaccination fails.

The early stages of an epidemic with vaccination is approximated by a branching process of (potentially) infected households, and the offspring of a given individual in the branching process are all households with which members of the single-household epidemic in the parent household make a potential global infectious contact.  As in the case without vaccination, the offspring distribution of this branching process is usually different in the initial generation from all subsequent generations.  Let $\tilde{C}'$ denote the offspring random variable for a non-initial individual.  Conditioning first on the size of the corresponding household and then on the number of people vaccinated in that household yields, in an obvious notation,
\begin{equation}
\label{fctildedash}
f_{\tilde{C}'} (s) = \sum_{n=1}^{\infty} \tilde{\rho}_n \sum_{v=0}^n x_{nv} f_{\tilde{C}_{n,v}'} (s).
\end{equation}
To determine $f_{\tilde{C}_{n,v}'} (s)$, consider a household in state $(n,v)$, i.e.\ of size $n$ having $v$ members vaccinated.  For $k=0,1,\ldots,v$, the probability that $k$ vaccinations are successful is $\binom{v}{k} \epsilon^k (1-\epsilon)^{v-k}$, and given that $k$ vaccinations are successful, the probability that the initial potentially contacted individual in that household is susceptible and thus triggers a local epidemic is $\frac{n-k}{n}$.  Moreover, if such a local epidemic is triggered, the number of potential infectious global contacts emanating from the local epidemic is distributed as $\tilde{C}^{(n-k)}$.  Thus,
\begin{equation}
\label{fctildedashnv}
f_{\tilde{C}_{n,v}'} (s) = \sum_{k=0}^{v} \binom{v}{k} \epsilon^k (1-\epsilon)^{v-k} \left[ \frac{(n-k)}{k} f_{\tilde{C}^{(n-k)}} (s) + \frac{k}{n} \right] .
\end{equation}

The distribution of the offspring random variable, $C'$ say, for the initial generation depends on how the initial infective for the network-households model is chosen.  For $n=1,2,\ldots$ and $v=0,1,\ldots,n$, let $p_{n,v}^V$ be the probability that a vaccinated individual chosen uniformly at random resides in a household in state $(n,v)$ and let $p_{n,v}^U$ be the corresponding probability for an unvaccinated individual.  Then
\begin{equation}
\label{pnvUV}
p_{n,v}^V = \frac{\tilde{\rho}_n x_{nv} \frac{v}{n}}{p_V} \quad \text{and} \quad
p_{n,v}^U = \frac{\tilde{\rho}_n x_{nv} \left( 1 - \frac{v}{n} \right)}{1-p_V}.
\end{equation}
Thus, if the epidemic is started by an individual chosen uniformly at random from all unvaccinated individuals being infected, then $C'$ is distributed as $C_U '$, where
\[
f_{C_U '} (s) = \sum_{n=1}^{\infty} \sum_{v=0}^{n-1} p_{n,v}^U \sum_{k=0}^v \binom{v}{k} \epsilon^k (1-\epsilon)^{v-k} f_{C^{(n-k)}} (s).
\]
Alternatively, if the epidemic is started by choosing a vaccinated individual uniformly at random, who triggers an outbreak only if its vaccination fails, then $C'$ is distributed as $C_V '$, where
\[
f_{C_V '} (s) = \epsilon + (1- \epsilon ) \sum_{n=1}^{\infty} \sum_{v=1}^n p_{n,v}^V \sum_{k=1}^v \binom{v-1}{k-1} \epsilon^{k-1} (1-\epsilon)^{v-k} f_{C^{(n-k)}} (s).
\]
The probability of a major outbreak may now be calculated as at the end of Section~\ref{subsecearlystages}. Again, the formulae simplify appreciably if the infectious period is constant.

A post-vaccination threshold parameter is given by $R_v = E[\tilde{C} ']$.  Let $f_{\tilde{C} '} ' (s)$ denote the first derivative of
$f_{\tilde{C}'} (s)$.  Then $R_v = f_{\tilde{C} '} ' (1)$, and differentiating (\ref{fctildedash}) and (\ref{fctildedashnv}) yields
\begin{equation}
\label{RvAoN}
R_v = \sum_{n=1}^{\infty} \tilde{\rho}_n \sum_{v=0}^n x_{nv} \mu_{n,v},
\end{equation}
where $\mu_{n,v} = E[ \tilde{C}_{n,v} ' ]$ is given by
\[
\mu_{n,v} = \sum_{k=0}^v \binom{v}{k} \epsilon^k (1-\epsilon)^{v-k} \left( \frac{n-k}{k} \right) E[ \tilde{C}^{(n-k)} ].
\]

As with the case of no vaccination, we can determine the relative final size of a major outbreak by considering a households-based branching process that approximates the susceptibility set of a typical individual.  As with the forward process, it is convenient to consider potential global neighbours when constructing this branching process.  Thus we start with an individual, $i^\ast$ say, chosen uniformly at random from the population, construct its local susceptibility set, taking the vaccine status of individuals in the household into account, then determine which global neighbours of individuals in this local susceptibility set would enter the susceptibility set of $i^{\ast}$ if they were susceptible (i.e.~unvaccinated or unsuccessfully vaccinated).  These individuals correspond to the first generation of the approximating backward branching process.  Suppose that there are $B'$ such individuals.  Next we take each of these $B'$ individuals in turn, first determine whether they really do enter the susceptibility set of $i^{\ast}$ (this happens with probability 1 if the individual is unvaccinated and with probability $1-\epsilon$ if it is vaccinated, independently for distinct individuals) and if they do enter the susceptibility set of $i^{\ast}$, determine the number of potential global neighbours of its local susceptibility set to obtain its offspring in the branching process, and so on.

Let $\tilde{B} '$ be the offspring random variable for any non-initial individual in this backward branching process.  Conditioning first on the state $(n,v)$ of that individual's household and then on whether it joins the susceptibility set of $i^{\ast}$, we obtain that
\[
f_{\tilde{B}'} (s) = \sum_{n=1}^{\infty} \tilde{\rho}_n \sum_{v=0}^n x_{nv} f_{\tilde{B}_{n,v}'} (s),
\]
where
\[
f_{\tilde{B}_{n,v}'} (s) = \sum_{k=0}^v \binom{v}{k} \epsilon^k (1-\epsilon)^{v-k} \left[ \frac{(n-k)}{k} f_{\tilde{B}^{(n-k)}} (s) + \frac{k}{n} \right].
\]

The distribution of $B'$ depends on how the initial individual $i^{\ast}$ is chosen.  If $i^{\ast}$ is chosen uniformly at random from all unvaccinated individuals, then $B'$ is distributed as $B_U '$, say, and conditioning on the state $(n,v)$ of $i^{\ast}$'s household yields
\[
f_{B_U '} (s) = \sum_{n=1}^{\infty} \sum_{v=0}^{n-1} p_{n,v}^U \sum_{k=0}^v \binom{v}{k} \epsilon^k (1-\epsilon)^{v-k} f_{B^{(n-k)}} (s),
\]
whilst if $i^{\ast}$ is chosen uniformly at random from all vaccinated individuals, then $B'$ is distributed as $B_V '$, where
\[
f_{B_V '} (s) = \sum_{n=1}^{\infty} \sum_{v=1}^n p_{n,v}^V \sum_{k=0}^v \binom{v}{k} \epsilon^k (1-\epsilon)^{v-k}
\left[ \frac{(n-k)}{n} f_{B^{(n-k)}} (s) + \frac{k}{n} \right] .
\]

Let $\xi^V$ be the smallest solution of $f_{\tilde{B} '} (s)=s$ in $[0,1]$.  Then the proportion of unvaccinated individuals that are ultimately infected by a major outbreak is $z^U = 1-f_{B_U '} ( \xi^V )$ and the corresponding proportion for vaccinated individuals is
$z^V = 1-f_{B_V '} (\xi^V )$.  The overall proportion of the population infected by a major outbreak is $z=p_V z^V + (1-p_V) z^U$.

\subsection{Non-random vaccine action}
\label{subsechousevacnr}

Analysing the consequences of a vaccination scheme using a non-random vaccine action model is more difficult than with an all-or-nothing vaccine action model since disease spread is now genuinely two type.  Thus we now consider two types of individual: type-$U$, unvaccinated, and type-$V$, vaccinated. The early stages of the epidemic can again be approximated by a branching process of infected households.  This is now a two-type branching process, with the type of an infected household being given by the type of the initial case in that household. The offspring of a given individual in the branching process correspond to the households that are contacted globally by members of that individual's corresponding single-household epidemic in the epidemic process.

Let $\bs{C}^U = (C_{UU}, C_{UV})$ denote the offspring random variable for the initial individual in the above branching process, given that individual is of type $U$, and let $\bs{C}^V = (C_{VU}, C_{VV})$ be the corresponding offspring random variable when the initial individual has type $V$.  Thus, for example, $C_{UU}$ and $C_{UV}$ are respectively the number of unvaccinated and vaccinated global infections that emanate from the initially infected household, given that the first infective in that household is unvaccinated.  Define $\tilde{\bs{C}}^U = ( \tilde{C}_{UU}, \tilde{C}_{UV})$ and $\tilde{\bs{C}}^V = ( \tilde{C}_{VU}, \tilde{C}_{VV})$ similarly for subsequent individuals in the branching process.  For
$\bs{s} = (s_U , s_V ) \in [0,1]^2$, let $f_{\bs{C}^U} (\bs{s}) = E[s_U^{C_{UU}} s_V^{C_{UV}} ]$ and define $f_{\bs{C}^V} (\bs{s})$,
$f_{\tilde{\bs{C}}^U} (\bs{s})$ and $f_{\tilde{\bs{C}}^V} (\bs{s})$ similarly.  (Here and henceforth, for any discrete random vector $\bs{X}$ we denote its joint PGF by $f_{\bs{X}}$.)  Recall that when the network of global contacts is formed half edges are paired uniformly at random.  It follows that, for $A \in \{ U,V \}$,
\[
f_{\tilde{\bs{C}}^A} (\bs{s})=\sum_{n=1}^{\infty} \sum_{v=0}^n p_{n,v}^A f_{\tilde{\bs{C}}_{n,v}^A} (\bs{s}),
\]
where $\tilde{\bs{C}}_{n,v}^A$ is a random vector giving the numbers of unvaccinated and vaccinated global infections that emanate from a non-initial infected household of size $n$, having $v$ members vaccinated, whose primary case is of type $A$.  The distribution of $\bs{C}^U$ and $\bs{C}^V$ depends on how the initial infective is chosen.  If it is chosen uniformly at random from all individuals of the appropriate type in the population, then, for $A \in \{ U,V \}$,
\[
f_{\bs{C}^A} (\bs{s}) = \sum_{n=1}^{\infty} \sum_{v=0}^n p_{n,v}^A f_{\bs{C}_{n,v}^A} (\bs{s}),
\]
where $\bs{C}_{n,v}^A$ is defined analogously to $\tilde{C}_{n,v}^A$ but for the initial infected household.

Let
\[
\tilde{M} = \left[ \begin{array}{ll}\tilde{m}_{UU}&\tilde{m}_{UV}\\\tilde{m}_{VU}&\tilde{m}_{VV}\end{array}\right],
\]
where, for example, $\tilde{m}_{UU} = E[ \tilde{C}_{UU} ]$.  The post-vaccination threshold parameter $R_v$ is given by the dominant eigenvalue (a real, positive eigenvalue of maximum modulus) of $\tilde{M}$.  It is well known (e.g.\ \citeA[p.\ 123]{HacJagVat2005}) that, provided $\tilde{m}_{UV} \tilde{m}_{VU} \neq 0$, the two-type branching process has non-zero probability of surviving if and only if $R_v > 1$.  Moreover, if $\tilde{m}_{UV} \tilde{m}_{VU} \neq 0$ and $R_v > 1$, then the survival probability (and hence the probability of a major outbreak) can be determined as follows.  Let $\bs{\sigma} = (\sigma_U, \sigma_V )$ be the unique solution in $[0,1)^2$ of the equations
\[
\sigma_U = f_{\tilde{C}^U} (\sigma_U, \sigma_V), \quad
\sigma_V = f_{\tilde{C}^V} (\sigma_U, \sigma_V).
\]
Then, if the epidemic is started by an unvaccinated individual chosen uniformly at random from the population becoming infected, the probability of a major outbreak is $\pmaj^U = 1-f_{C^U} (\bs{\sigma})$.  The corresponding probability when the initial infective is a vaccinated individual is
$\pmaj^V = 1 - f_{C^V} (\bs{\sigma})$. Calculation of the PGFs $f_{\bs{C}^U_{n,v}}$, $f_{\bs{C}^V_{n,v}}$, $f_{\tilde{\bs{C}}^U_{n,v}}$ and $f_{\tilde{\bs{C}}^V_{n,v}}$, which is rather involved unless the infectious period is constant, is described in Appendix~\ref{AppHouseVac}.  Calculation of $\tilde{M}$, which is sufficient for determining optimal vaccination strategies, is simpler and we now outline it.

Consider first a local (single-household) epidemic in a household in state $(n,v)$, initiated by one of the household members becoming infected.  Let $T_U^{(n,v)}$ and $T_V^{(n,v)}$ denote the number of unvaccinated and vaccinated individuals ultimately infected by this local epidemic, not including the initial case.  For $A,A' \in \{ U,V \}$, let $\mu^{(n,v)} (A,A')$ be the mean of $T_{A'}^{(n,v)}$ given that the initial case is of type $A$.  (Calculation of $\mu^{(n,v)} (A,A')$ is described in Appendix~\ref{AppMeanLocalFS}.) Also, define the marginal transmission probabilities $p_G^{NR} (U,U)$, $p_G^{NR} (U,V)$, $p_G^{NR} (V,U)$ and $p_G^{NR} (V,V)$ between unvaccinated and vaccinated global neighbours, where, for example, $p_G^{NR} (U,V)$ is the probability that an unvaccinated infective infects a given vaccinated global neighbour.  Then
\begin{equation}
\label{PGNR}
P_G^{NR} = \left[
\begin{array}{ll}
p_G^{NR} (U,U)&p_G^{NR} (U,V)\\
p_G^{NR} (U,V)&p_G^{NR} (V,V)\end{array}\right] = \left[
\begin{array}{ll}1-\phi_I (\lambda_G)&1-\phi_I (a \lambda_G)\\
1-\phi_I (b \lambda_G)&1-\phi_I (ab \lambda_G )\end{array}\right] ,
\end{equation}
where $\phi_I (\theta) = E[\re^{-\theta I} ]$ ($\theta \geq 0$) is the moment generating function (MGF) of the infectious period random variable $I$.

Taking expectations with respect to the state $(n,v)$ of the infected household shows that, for $A,A' \in \{ U,V\}$,
\begin{equation}
\tilde{m}_{AA'} = \sum_{n=1}^{\infty} \sum_{v=0}^n p_{n,v}^A \tilde{m}_{AA'}^{(n,v)}, \label{mtilde}
\end{equation}
where $\tilde{m}_{AA'}^{(n,v)}$ is defined analogously to $\tilde{m}_{AA'}$ but for a household in state $(n,v)$.  Further, arguing as in the derivations of (\ref{ECtilden}) and (\ref{Rstar}), yields
\begin{equation}
\tilde{m}_{AA'}^{(nv)} = \left[ \mu_{\tilde{D}-1} p_G^{NR} (A,A') + \mu_D \left( \mu^{(n,v)} (A,U) p_G^{NR} (U,A')
+ \mu^{(n,v)} (A,V) p_G^{NR} (V,A')\right)\right]p_{A'}, \label{mtildenv}
\end{equation}
where $p_U = 1-p_V$.  Hence, if we let
\[
F=\left[ \begin{array}{ll}F_{UU}&F_{UV}\\F_{VU}&F_{VV}\end{array}\right] \quad \text{and}\quad D_V=
\left[ \begin{array}{cc}1-p_V&0\\0&p_V\end{array}\right],
\]
where, for $A,A' \in \{U,V\}$,
\[
F_{AA'}=\sum_{n=1}^{\infty} \sum_{v=0}^n p_{n,v}^A \mu^{(n,v)}(A,A'),
\]
then (\ref{mtilde}) and (\ref{mtildenv}) yield
\begin{equation}
\tilde{M}=(\mu_{\tilde{D} -1} I + \mu_D F) P_G^{NR} D_V .\label{Mtilde}
\end{equation}

Turning to the relative final size of a major outbreak, we consider a households-based branching process that approximates the susceptibility set of a given individual.  This is now a two-type branching process, with type ($U$ or $V$) corresponding to the type of the primary member of the corresponding local susceptibility set.  Define the offspring random variables
$\bs{B}^U = (B_{UU}, B_{UV})$,
$\bs{B}^V = (B_{VU}, B_{VV})$,
$\tilde{\bs{B}}^U = (\tilde{B}_{UU}, \tilde{B}_{UV})$ and
$\tilde{\bs{B}}^V = ( \tilde{B}_{VU}, \tilde{B}_{VV})$
for this branching process in the obvious fashion (cf.\ the forward process offspring random variables $\bs{C}^U$, $\bs{C}^V$, $\tilde{\bs{C}}^U$ and $\tilde{\bs{C}}^V$ and the notation used in Section~\ref{subsechousevacaon}).  We determine first the PGFs $f_{\tilde{\bs{B}}^U} (\bs{s})$ and $f_{\tilde{\bs{B}}^V} (\bs{s})$.

First note that, for $A \in \{ U,V \}$, conditioning on the state $(n,v)$ of a household yields, in obvious notation,
\[
f_{\tilde{\bs{B}}^A} (\bs{s}) = \sum_{n=1}^\infty \sum_{v=0}^n p_{n,v}^A f_{\tilde{\bs{B}}_{n,v}^A} (\bs{s}).
\]
Fix $A \in \{ U,V \}$ and $(n,v)$, and let $\bs{M}_A^{(n,v)} = (M_{AU}^{(n,v)}, M_{AV}^{(n,v)})$, where $M_{AU}^{(n,v)}$ and
$M_{AV}^{(n,v)}$ are respectively the numbers of unvaccinated and vaccinated individuals, not including $i^{\ast}$ itself, in the local susceptibility set of a typical type-$A$ individual, $i^{\ast}$ say, that resides in a household in state $(n,v)$.
Then $\tilde{\bs{B}}_{n,v}^A$ admits the decomposition
\begin{equation}
\tilde{\bs{B}}_{n,v}^A = \tilde{\bs{B}}_{n,v}^{A0} + \sum_{i=1}^{M_{AU}^{(n,v)}} \tilde{\bs{B}}_{n,v}^{AV} (i) +
\sum_{j=1}^{M_{AV}^{(n,v)}} \tilde{\bs{B}}_{n,v}^{AV} (j), \label{BtildeA}
\end{equation}
where $\tilde{\bs{B}}_{n,v}^{A0}$, $\tilde{\bs{B}}_{n,v}^{AU} (i)$ and $\tilde{\bs{B}}_{n,v}^{AV} (j)$ are the contributions to $\tilde{\bs{B}}_{n,v}^A$ from the primary individual $i^{\ast}$, the $i$th secondary unvaccinated member and the $j$th secondary vaccinated member of $i^{\ast}$'s local susceptibility set, respectively.  Let $D_{i^\ast}$ denote the number of neighbours $i^{\ast}$ has in the global network, so $D_{i^\ast} \sim \tilde{D}$, and recall that one of these global neighbours is used when $i^\ast$ joins the susceptibility set process.  Thus,
\[
\bs{B}_{n,V}^{A0} = \sum_{k=1}^{D_{i^\ast}-1} \bs{\chi}_k^A,
\]
where $\bs{\chi}_k^A = (1,0)$ if the $k$th global neighbour of $i^\ast$ is unvaccinated and joins the susceptibility set process,
$\bs{\chi}_k^A = (0,1)$ if this neighbour is vaccinated and joins the susceptibility set process, and $\bs{\chi}_k^A = (0,0)$ otherwise.  Note that, independently, each such global neighbour is vaccinated with probability $p_V$, and it joins the susceptibility set process with probability
$p_G^{NR} (U,A)$ if it is unvaccinated and with probability $p_G^{NR} (V,A)$ if it is vaccinated.  Thus
\[
f_{\bs{B}_{n,V}^{A0}} (\bs{s}) = f_{\tilde{D}-1} (p^A (\bs{s})),
\]
where
\begin{align*}
p^A (\bs{s}) &= f_{\bs{\chi}_1^A} (\bs{s})\\
&=(1-p_V) p_G^{NR} (U,A) s_U + p_V p_G^{NR} (V,A) s_V
 + 1 - (1-p_V) p_G^{NR} (U,A) - p_V p_G^{NR} (V,A).
\end{align*}
A similar argument shows that $f_{\tilde{\bs{B}}_{n,v}^{AU} (i)} (\bs{s})=f_D (p^U (\bs{s}))$ and
$f_{\tilde{\bs{B}}_{n,v}^{AV} (j)} (\bs{s}) = f_D (p^V (\bs{s}))$, and exploiting the mutual independence of all random quantities in (\ref{BtildeA}) except the components of $\bs{M}_A^{(n,v)}$ then yields
\[
f_{\tilde{\bs{B}}_{n,v}^A} (\bs{s}) = f_{\tilde{D}-1} (p^A (\bs{s})) f_{\bs{M}_A^{(n,v)}} (f_D (p^U (\bs{s})), f_D(p^V (\bs{s}))).
\]

The distribution of $\bs{B}^U$ and $\bs{B}^V$ depend on how the initial individual for the susceptibility set process is chosen.
For $A \in \{ U,V \}$, if this initial individual is chosen uniformly at random from all type-$A$ individuals in the population, then
\[
f_{\bs{B}^A} (\bs{s}) = \sum_{n=1}^\infty \sum_{v=0}^n p_{n,v}^A f_{\bs{B}_{n,v}^A} (\bs{s}),
\]
where
\[
f_{\bs{B}_{n,v}^A} (\bs{s}) = f_D (p^A (\bs{s})) f_{\bs{M}_A^{(n,v)}} (f_D (p^U (\bs{s})), f_D (p^V (\bs{s}))).
\]
Suppose that $\tilde{m}_{UV} \tilde{m}_{VU} \neq 0$ and $R_v > 1$.  Let $\bs{\xi} = ( \xi_U, \xi_V )$ be the unique solution in $[0,1)^2$ of the equations
\[
\xi_U = f_{\tilde{\bs{B}}^U} ( \xi_U, \xi_V ), \quad
\xi_V = f_{\tilde{\bs{B}}^V} ( \xi_U, \xi_V ).
\]
Then the proportions of unvaccinated and vaccinated individuals that are infected by a major epidemic are given by
$z^U = f_{\bs{B}^U} (\xi_U, \xi_V)$ and $z^V = f_{\bs{B}^V} (\xi_U, \xi_V)$, respectively.

\subsection{Optimal vaccination strategies} 
\label{subsecopt}

A main aim of a vaccination scheme is to reduce the threshold parameter $R_{\ast}$ to below one, i.e.\ to make $R_v \leq 1$, and thus prevent a major outbreak occurring.  The vaccine response may be such that $R_v > 1$ even if the entire population is vaccinated, in which case vaccination by itself is insufficient to be sure of preventing a major outbreak.  However, if $R_{\ast} > 1$ and it is possible to make $R_v \leq 1$ then it is of interest to determine the allocation of vaccines that reduces $R_v$ to 1 with the minimum vaccination coverage $p_V$.

Suppose that the population has a maximum household size $n_{\max} < \infty$.  Then $p_V$ is a linear function of $x_{nv}$ ($n=1,2,\ldots$,
$n_{\max}$; $v=0,1,\ldots,n$) (recall (\ref{pv})), as is $R_v$ when the vaccine action is all-or-nothing (recall (\ref{RvAoN})).  Thus in this case determining the allocation of vaccines that (a) minimises $p_V$ subject to $R_v \leq 1$ or (b) minimises $R_v$ subject to an upper bound on $p_V$ are both linear programming problems.  Moreover, as we outline below,
the method of \citeA{BalLyn2002,BalLyn2006} can be used to construct explicitly the solutions of these linear programming problems.  The situation is in general more complicated if the vaccine action is non-random, since then $R_v$ is then the dominant eigenvalue of the matrix $\tilde{M}$, and the corresponding optimisation problems are non-linear.  However, the problem is linear if $\textrm{rank} (\tilde{M})=1$, a sufficient condition for which is $\textrm{rank} (P_G^{NR})=1$, i.e.\ $p_G^{NR} (U,U) p_G^{NR} (V,V) = p_G^{NR} (U,V) p_G^{NR} (V,U)$.  Note that $\textrm{rank} (P_G^{NR})=1$ if either $a=1$ or $b=1$, so a leaky vaccine response satisfies this constraint.

Consider the non-random vaccine response and suppose that $\textrm{rank} (\tilde{M})=1$.  Then $R_v = \textrm{trace} (\tilde{M})$ and, recalling (\ref{Mtilde}), it follows using (\ref{pnvUV}), (\ref{mtilde}) and (\ref{mtildenv}) that
\begin{equation}
R_v = \sum_{n=1}^\infty \sum_{v=0}^n \tilde{\rho}_n x_{nv} \mu_{n,v}^{NR}, \label{RvNR}
\end{equation}
where
\begin{align}
\mu_{n,v}^{NR} &= \left( 1 - \frac{v}{n} \right) \left\{ \tilde{\mu}_{D-1} p_G^{NR} (U,U) + \mu_D \left[ \mu^{(n,v)} (U,U) p_G^{NR} (U,U) + \mu^{(n,v)}
(U,V) p_G^{NR} (V,U)\right] \right\}\notag\\
&\qquad + \frac{v}{n} \left\{ \tilde{\mu}_{D-1} p_G^{NR} (V,V) + \mu_D \left[ \mu^{(n,v)} (V,U) p_G^{NR} (U,V) + \mu^{(n,v)} (V,V)
p_G^{NR} (V,V)\right]\right\}.\label{MUNRnv}
\end{align}
Observe that, when $\textrm{rank} (\tilde{M})=1$, $R_v$ takes the same form as for the all-or-nothing vaccine response; compare (\ref{RvNR}) and (\ref{RvAoN}).

To characterise the optimal vaccination schemes in these cases, it is convenient to consider a finite population of $m$ households, with maximum household size $n_{\max}$. Let $m_n = m \rho_n$ be the number of households of size $n$ and let $h_{nv} = m_n x_{nv}$ be the number of households in state $(n,v)$.  Then $\tilde{\rho}_n = nm_n / N$, where $N$ is the total population size, and, writing $\mu_{n,v}$ for $\mu_{n,v}^{A o N}$ or
$\mu_{n,v}^{NR}$, as appropriate, (\ref{RvAoN}) or (\ref{RvNR}) implies that
\begin{equation}
R_v = \sum_{n=1}^{n_{\max}} \sum_{v=0}^n h_{nv} M_{n,v},\label{Rv1}
\end{equation}
where $M_{n,v}=n \mu_{n,v} / N$, and (\ref{pv}) yields
\begin{equation}
p_V = \frac{1}{N} \sum_{n=1}^{n_{\max}} \sum_{v=0}^n vh_{nv}.\label{pv1}
\end{equation}

Observe that (\ref{Rv1}) implies that $R_v$ is obtained by summing $M_{n,v}$ over all households in the population.  For $n=1,2,\ldots,n_{\max}$ and $v=0,1,\ldots,n-1$, let $G_{n,v}=M_{n,v} - M_{n,v+1}$ be the reduction in $R_v$ obtained by vaccinating one further individual in a household in state $(n,v)$.  If $G_{n,v}$ is decreasing in $v$ for each fixed $n$ (so successive vaccinations in the same household have diminishing returns), then it is straightforward to determine optimal vaccination schemes \cite{BalLyn2002,BalLyn2006}.  One simply orders the states $(n,v)$ according to decreasing $G_{n,v}$, and then uses this ordering to determine the order in which individuals in the population are vaccinated, stopping the process when either the vaccination coverage reaches the desired level or when $R_v \leq 1$, depending on the optimisation problem under consideration. (The `worst' scheme is obtained by vaccinating whole households in increasing order of $M_{n,0}-M_{n,n}$.)
If for some $n$, say $n=n'$, $G_{n',v}$ is not decreasing in $v$, then only those states, $(n',v')$ say, on the lower edge of the convex hull of
$\{ (v,G_{n',v}) \colon v=0,1,\ldots,n'-1\}$ can be part of an optimal vaccination scheme.  It is still possible to give explicit solutions of associated optimisation problems (cf.\ \citeA{BalBriLyn2004}), and of `worst' schemes, but the details are more involved.

%%%%%%%%%%%%%%%%%%%% SECTION FOUR

\section{Acquaintance vaccination}
\label{secacq}

\subsection{Introduction}
\label{subsecacqintro}

The acquaintance vaccine allocation model proposed by \citeA{CohHavben2003} and further analysed by \citeA{BriJanMar2007}, both in the setting of a population modeled by the configuration model (without household structure), is as follows.  Each individual in the population is sampled independently a Poisson distributed number of times, with mean $\kappa>0$, and each time an individual is sampled it chooses one of its neighbours uniformly at random, with replacement, and that neighbour is vaccinated.  If a sampled individual has no neighbours then that sampling is ignored.  Individuals are vaccinated at most once, even if they are chosen to be vaccinated more than once.

If the vaccine is perfect then the early stages and final outcome of an epidemic with this acquaintance vaccination model can be analysed relatively easily using branching process approximations.  Essentially this is because the epidemic involves only unvaccinated individuals, and the degrees of the neighbours of an unvaccinated individual are mutually independent.  However, if the vaccine is imperfect, then the epidemic may also involve vaccinated individuals and the degrees of the neighbours of a vaccinated individual are dependent.  (A low-degree neighbour of a given individual, $A$ say, is more likely to nominate $A$ for vaccination than a high-degree neighbour but, if $A$ is vaccinated, at least one of $A$'s neighbours nominated $A$ for vaccination.)  It follows that the independence property required for a branching process approximation breaks down.  This dependence can be overcome if individuals are also typed by their degree but, unless the support of $D$ is small, the calculations become computationally prohibitively expensive. Indeed, infinite-type branching processes are required if the support of $D$ is countably infinite.  For these reasons, \citeA{BalSir2013} introduced an alternative acquaintance vaccine allocation model and analysed it in the setting of a standard network model (i.e.\ without households).  We now extend this analysis to the network-households model.

\subsection{Model and preliminary results}
\label{subsecacqmod}

We assume that each individual is sampled independently with probability $p_S$ and then each sampled individual nominates each of its global neighbours independently with probability $p_N$.  All individuals that are nominated at least once are then vaccinated.  Thus individuals are sampled only once and it is easily seen that the degrees of the neighbours of both vaccinated and unvaccinated individuals are mutually independent, thus facilitating branching process approximations which do not involve typing by degree.

We approximate the early stages of the epidemic, with vaccination, by a multitype branching process of infected households, in which households are typed by the type of their primary (globally contacted) case.  Such primary cases are typed by (i) whether they are named ($N$), vaccinated ($V$) or unvaccinated ($U$) and (ii) whether or not they are sampled and thus might name their neighbour for vaccination ($S$ and $S^c$).  Here $N$ means that a primary case was named by its global infector, and therefore is vaccinated; $V$ means that it is not named by its global infector but it is nevertheless vaccinated (i.e.\ it is named by another neighbour) and $U$ means that it is unvaccinated (i.e.\ not named by any of its neighbours).  Thus there are 6 types of infected households, which for notational convenience we give the following numerical indices:
\[
\begin{array}{cccccc}
1&2&3&4&5&6\medskip\\
(N,S)&(V,S)&(U,S)&(N,S^c)&(V,S^c)&(U,S^c)
\end{array}
\]
Secondary infected cases in a household, and also the primary case in the initially infected household, need only to be typed $V$ or $U$, according to whether or not they are vaccinated.  A similar typing is used for the backward process.  In Section~\ref{subsecacqthreshold} we determine the mean offspring matrix for the forward process, and hence the post-vaccination threshold parameter $R_v$, and in Section~\ref{subsecacqpmaj} we determine the offspring distribution PGFs for the backward process, and hence the relative final size of a major epidemic.  The offspring distribution PGFs for the forward process are generally more complicated and their calculation is described in Appendix~\ref{AppAcqVac}.  We first derive some elementary properties pertaining to this acquaintance vaccination model.

First, note that the probability that an individual is not named by a given neighbour is $1-p_S p_N$, so the probability it is vaccinated is
\begin{equation}
p_V = 1 - \sum_{d=0}^{\infty} p_d (1-p_S p_N)^d = 1-f_D (1-p_S p_N ),
\label{pV}
\end{equation}
which, of course, also gives the vaccination coverage.  Let $D_V$ and $D_U$ denote the degree of a typical vaccinated and unvaccinated individual, respectively.  Then,
\begin{equation}
P(D_U =d) = \frac{P(D=d)P(U|D=d)}{P(U)}=\frac{p_d (1-p_S p_N)^d}{1-p_V} \quad (d=0,1,\ldots)\label{pDUd}
\end{equation}
and, similarly,
\begin{equation}
P(D_V=d)=\frac{p_d (1-(1-p_S p_N)^d)}{p_V} \quad (d=1,2,\ldots).\label{pDVd}
\end{equation}

Note that whether or not an individual is sampled is independent of its degree.  Thus primary cases of types 1 and 4 have the same degree distribution, as do primary cases of types 2 and 5, and primary cases of types 3 and 6.  Let $\tilde{D}_N$, $\tilde{D}_V$ and $\tilde{D}_U$ denote generic random variables having these respective distributions.  First note that $\tilde{D}_N \stackrel{D}{=} \tilde{D}$.  Second, consider a typical unvaccinated primary case.  It has unconditional degree distribution $\tilde{D}$ but, in addition to not being named by its infector, we also know that it is not named by any of its other global neighbours.  Thus
\begin{equation}
P(\tilde{D}_U = d) = \frac{P(\tilde{D}=d)P(U|\tilde{D}=d)}{1-\tilde{p}_V} =
\frac{\tilde{p}_d (1-p_N p_S)^{d-1}}{1-\tilde{p}_V} \quad (d=1,2,\ldots),\label{pdtildeu}
\end{equation}
where
\begin{equation}
\tilde{p}_V = \sum_{d=1}^{\infty} \tilde{p}_d (1-(1-p_N p_S)^{d-1}) = 1-f_{\tilde{D}-1} (1-p_N p_S)
\label{pVtilde}
\end{equation}
is the probability that a typical unnamed neighbour of an infector is vaccinated.  Similarly,
\begin{equation}
P(\tilde{D}_V=d)=\frac{\tilde{p}_d (1-(1-p_N p_S)^{d-1})}{\tilde{p}_V} \quad (d=2,3,\ldots).\label{pdtildev}
\end{equation}

\subsection{Threshold parameter}
\label{subsecacqthreshold}

As far as is possible, we treat the all-or-nothing and non-random models simultaneously.  In contrast to Section~\ref{subsechousevacaon}, when analysing the all-or-nothing vaccine response, we consider actual, rather than potential, global infections.  The forward branching process for both models is then similar to that used for the non-random model in Section~\ref{subsechousevacnr}, except now there are 6 types of individuals.  As always, the offspring distributions are different in the initial generation from subsequent generations, but only those for the latter are required to determine the threshold parameter.  For $i=1,2,\ldots,6$, let $\tilde{\bs{C}}_i = (\tilde{C}_{i1}, \tilde{C}_{i2}, \ldots, \tilde{C}_{i6})$ denote the offspring random variable for a type-$i$ non-initial individual in the forward branching process.  Thus $\tilde{C}_{ij}$ is the number of type-$j$ primary household cases emanating from a typical single household epidemic that is initiated by a single type-$i$ primary case.  Our goal is to determine the matrix $\tilde{M}=[\tilde{m}_{ij}]$, where $\tilde{m}_{ij}=E[\tilde{C}_{ij}]$; recall that $R_v$ is the dominant eigenvalue of $\tilde{M}$.

To determine $\tilde{m}_{ij}$ it is convenient to decompose $\tilde{m}_{ij}$ into
\[
\tilde{m}_{ij} = \tilde{m}_{ij}^P + \tilde{m}_{ij}^S ,
\]
where $\tilde{m}_{ij}^P$ is the mean number of type-$j$ primary household cases emanating from the primary type-$i$ case and $\tilde{m}_{ij}^S$ is the mean total number of primary type-$j$ household cases emanating from all secondary cases in the single-household epidemic.  %We consider first $\tilde{m}_{ij}^P$.

\subsubsection*{Network infections emamating from primary cases}
Define the matrix $P_G$ of marginal global transmission probabilities, so $P_G = P_G^{NR}$ (recall (\ref{PGNR})) if the vaccine action is non-random and $P_G = P_G^{AoN}$ if the vaccine action is all-or-nothing, where
\[
P_G^{AoN} = \left[ \begin{array}{ll}
p_G^{AoN} (U,U)&p_G^{AoN} (U,V)\\
p_G^{AoN} (V,U)&p_G^{AoN} (V,V)\end{array}\right] = p_G
\left[ \begin{array}{ll}1&1-\epsilon\\1&1-\epsilon \end{array}\right].
\]
The marginal global transmission probabilities, for the types $i,j = 1,2,\ldots,6$, are then given by $p_{ij}^G = p_G (A(i),A(j))$, where
$A(k) = U$ if $k=3,6$ and $A(k)=V$ if $k=1,2,4,5$.  It follows that
\[
\tilde{m}_{ij}^P = \tilde{\mu}_i^G \hat{p}_{ij} p_{ij}^G ,
\]
where $\tilde{\mu}_i^G$ is the mean number of global neighbours a typical type-$i$ primary case has in the forward process (i.e.\ ignoring the infector of this primary case) and $\hat{p}_{ij}$ is the probability that a given such global neighbour is of type $j$.  We now determine $\tilde{\mu}_i^G$ $(i=1,2,\ldots,6)$ and $\hat{p}_{ij}$ $(i,j=1,2,\ldots,6)$.

Consider a type-1 (i.e.\ $(N,S)$) primary case.  Its global degree is distributed as $\tilde{D}_N \stackrel{D}{=} \tilde{D}$, so
$\tilde{\mu}_1^G = \mu_{\tilde{D}-1}$.  A given neighbour is named with probability $p_N$, and, if that neighbour is unnamed, it is vaccinated with probability $\tilde{p}_V$ and otherwise unvaccinated.  Further, independently, that neighbour is sampled with probability $p_S$.  Thus,
$\hat{p}_{11} = p_N p_S$, $\hat{p}_{12} = (1-p_N) \tilde{p}_V p_S$, $\hat{p}_{13} = (1-p_N)(1-\tilde{p}_V)p_S$, $\hat{p}_{14}=p_N (1-p_S)$,
$\hat{p}_{15} = (1-p_N) \tilde{p}_V (1-p_S)$ and $\hat{p}_{16} = (1-p_N) (1-\tilde{p}_V )(1-p_S)$.  The situation is similar for a type-4 (i.e.\ $(N,S^c)$) individual, except a type-4 individual is not sampled and hence cannot name its global neighbours.  Hence $\tilde{\mu}_4^G = \mu_{\tilde{D}-1}$, $\hat{p}_{41} = \hat{p}_{44}=0$, $\hat{p}_{42} = \tilde{p}_V p_S$,
$\hat{p}_{43} = (1-\tilde{p}_V) p_S$, $\hat{p}_{45} = \tilde{p}_V (1-p_S)$ and $\hat{p}_{46} = (1-\tilde{p}_V)(1-p_S)$.

Next consider a type-2 (i.e.\ $(V,S)$) primary case, $i^{\ast}$ say.  Its global degree is distributed as $\tilde{D}_V$, so
$\tilde{\mu}_2^G = \mu_{\tilde{D}_V -1}$ and, using (\ref{pdtildev}),
\[
\mu_{\tilde{D}_V -1} = \frac{\mu_{\tilde{D}-1} - (1-p_N p_S) f'_{\tilde{D}-1} (1-p_N p_S)}{\tilde{p}_V},
\]
Let $\tilde{N}_S$ be the number of susceptible global neighbours of $i^{\ast}$ that are sampled.  Then $\tilde{N}_S \geq 1$, since $i^{\ast}$ is vaccinated but not named by its global infector, and
\begin{equation}
P(\tilde{N}_S = k \mid \tilde{D}_V = d) = \frac{\binom{d-1}{k} p_S^k (1-p_S)^{d-1-k} (1-(1-p_N)^k)}{1-(1-p_N p_S)^{d-1}} \quad (k=1,2,\ldots,d-1),
\label{ptildensdv}
\end{equation}
whence, using (\ref{pdtildev}),
\begin{equation}
\mu_{\tilde{N}_S}= \frac{p_S \left(\mu_{\tilde{D}-1} - (1-p_N) f'_{\tilde{D}-1}(1-p_N p_S )\right)}{\tilde{p}_V}.
\label{muNtildeS}
\end{equation}
It follows that the probability that a given susceptible neighbour of $i^{\ast}$ is sampled is $p_S^{\tilde{V}} = \mu_{\tilde{N}_S} / \tilde{\mu}_2^G$ and that, for $j=1,2,\ldots,6$, $\hat{p}_{2j}$ is given by $\hat{p}_{1j}$ with $p_S$ replaced by $p_S^{\tilde{V}}$.  Similar arguments for a type-5 (i.e.\ $(U,S^c)$) primary case yield $\tilde{\mu}_5^G = \mu_{\tilde{D}_V -1}$ and, for $j=1,2,\ldots,6$, $\hat{p}_{5j}$ is given by $\hat{p}_{4j}$ with $p_S$ replaced by $p_S^{\tilde{V}}$.

Now consider a type-3 (i.e.\ $(U,S)$) primary case, $j^{\ast}$ say.  Its global degree is distributed as $\tilde{D}_U$, so
$\tilde{\mu}_3^G = \mu_{\tilde{D}_U-1}$ and, using (\ref{pdtildeu}),
\begin{equation}
\mu_{\tilde{D}_U-1} = \frac{(1-p_N p_S) f'_{\tilde{D}-1} (1-p_N p_S)}{1-\tilde{p}_V}.
\label{muDUtilde1}
\end{equation}
Since $j^{\ast}$ is not vaccinated, none of its neighbours name $j^{\ast}$, so the probability that a given neighbour, $k^{\ast}$ say, is sampled is given by
\[
p_S^U = P(k^{\ast} \text{ sampled} \mid k^{\ast} \text{ does not name } j^\ast )=\frac{p_S (1-p_N)}{1-p_S p_N}.
\]
It follows that, for $j=1,2,\ldots,6$, $\hat{p}_{3j}$ is given by $\hat{p}_{1j}$ with $p_S$ replaced by $p_S^U$.  Similarly,
$\tilde{\mu}_6^G = \mu_{\tilde{D}_U-1}$ and, for $j=1,2,\ldots,6$, $\hat{p}_{6j}$ is given by $\hat{p}_{4j}$ with $p_S$ replaced by $p_S^U$.

\subsubsection*{Network infections emamating from secondary cases}
%Now consider $\tilde{m}_{ij}^S$ and note first that, 
First we note that, 
for $j=1,2,\ldots,6$, $\tilde{m}_{1j}^S = \tilde{m}_{2j}^S = \tilde{m}_{4j}^S = \tilde{m}_{5j}^S$ ($=\tilde{m}_{Vj}$ say) and $\tilde{m}_{3j}^S = \tilde{m}_{6j}^S$ ($=\tilde{m}_{Uj}^S$ say).  Further, for $A \in \{ U,V \}$ and $j=1,2,\ldots,6$,
\[
\tilde{m}_{Aj}^S = \mu_S (A,U) \hat{m}_{Uj} + \mu_S (A,V) \hat{m}_{Vj},
\]
where, for $A,A' \in \{U,V\}$, $\mu_S (A,A')$ is the mean number of type-$A'$ secondary cases in a typical single-household epidemic initiated by a primary case of type $A$ and, for $A \in \{ U,V \}$, $\hat{m}_{Aj}$ is the mean number of type-$j$ primary cases emanating from a typical type-$A$ secondary case.

Conditioning on the size of a typical globally contacted household, we obtain in an obvious notation that
\[
\mu_S (A,A') = \sum_{n=1}^{\infty} \tilde{\rho}_n \mu_S^{(n)} (A,A') \quad (A,A' \in \{U,V\}).
\]
Further, all secondary individuals in the household are vaccinated independently, each with probability $p_V$, so, again in an obvious notation,
\[
\mu_S^{(n)} (A,A') = \sum_{k=0}^{n-1} \binom{n-1}{k} p_V^k (1-p_V)^{n-1-k} \mu_S^{(n,k)} (A,A').
\]
If the vaccine action is non-random then, recalling the notation in Section~\ref{subsechousevacnr}, $\mu_S^{(n,k)} (A,A') = \mu_{NR}^{(n,k+\delta_{A,V})} (A,A')$, where $\delta_{A,V}=1$ if $A=V$ and 0 otherwise.  If the vaccine action is all-or-nothing, we condition on the number of successfully vaccinated secondary individuals in the household to obtain, for $A \in \{ U,V \}$,
\[
\mu_S^{(n,k)} (A,V) = \sum_{l=0}^{\min(k,n-2)} \binom{k}{l} \epsilon^l (1-\epsilon) ^{k-l} \mu^{(n-l)} (\lambda_L) \left( \frac{k-l}{n-l-1}\right)
\]
and
\[
\mu_S^{(n,k)} (A,U) = \sum_{l=0}^{\min(k,n-2)} \binom{k}{l} \epsilon^l (1-\epsilon)^{k-l} \mu^{(n-l)} (\lambda_L) \left( \frac{n-k-1}{n-l-1} \right).
\]
(Note that $\mu^{(n-1)}(\lambda_L)=0$ when $k=l=n-1$.)

To determine $\hat{m}_{Uj}$, note that if a secondary case, $i^{\ast}$ say, is unvaccinated then its degree is distributed according to $D_U$ and, using (\ref{pDUd}), its expected number of global neighbours is
\begin{equation}
\mu_{D_U} = (1-p_S p_N) f'_D (1-p_S p_N)/(1-p_V).
\label{muDU}
\end{equation}
Further, $i^\ast$ is sampled with probability $p_S$, and the type of a given global neighbour of $i^{\ast}$ is distributed according to $\hat{p}_{3j}$ if $i^{\ast}$ is sampled and according to $\hat{p}_{6j}$ if $i^{\ast}$ is not sampled.  Thus,
\[
\hat{m}_{Uj} = \mu_{D_U} \left(p_S \hat{p}_{3j} p_{3j}^G + (1-p_S) \hat{p}_{6j} p_{6j}^G \right) \quad (j=1,2,\ldots,6).
\]

Finally, to derive $\hat{m}_{Vj}$, let $N_S$ and $N_{S^c}$ denote the number of global neighbours of a typical vaccinated secondary case, $j^{\ast}$ say, that are sampled and not sampled, respectively.  Then arguing as in the derivation of (\ref{muNtildeS}) shows that
\[
\mu_{N_S} = \frac{p_S \left(\mu_D - (1-p_N) f'_D (1-p_S p_N)\right)}{p_V}
\]
and
\[
\mu_{N_{S^c}} = \frac{(1-p_S)\left(\mu_D - f'_D (1-p_S p_N)\right)}{p_V}.
\]
Further, $j^{\ast}$ is sampled with probability $p_S$. If $j^{\ast}$ is sampled then a given neighbour is named, vaccinated and unvaccinated with probability $p_N$, $(1-p_N) \tilde{p}_V$ and $(1-p_N)(1-\tilde{p}_V)$, respectively; whilst if $j^{\ast}$ is not sampled these probabilities are 0, $\tilde{p}_V$ and $1-\tilde{p}_V$. We therefore have
\begin{equation*}
\hat{m}_{Vj} = \mu_{D_V}  \left(p_S \hat{p}_{1j} p_{1j}^G + (1-p_S) \hat{p}_{4j} p_{4j}^G \right) \quad (j=1,2,\ldots,6).
\end{equation*}
%$\hat{m}_{V1} = \mu_{N_S} p_S p_N p^G_{11}$,
%$\hat{m}_{V2} = \mu_{N_S} p_S (1-p_N) \tilde{p}_V p^G_{12}$,
%$\hat{m}_{V3} = \mu_{N_S} p_S (1-p_N) (1-\tilde{p}_V) p^G_{13}$,
%$\hat{m}_{V4} = 0$,
%$\hat{m}_{V5} = \mu_{N_{S^c}} (1-p_S) \tilde{p}_V p^G_{15}$ and
%$\hat{m}_{V6} = \mu_{N_{S^c}} (1-p_S)(1-\tilde{p}_V) p^G_{16}$.

\subsection{Probability of a major outbreak}
\label{subsecacqpmaj}

To determine the probability of a major outbreak we need to derive the PGFs $f_{\tilde{C}_i} (\bs{s})$, where $\bs{s}=(s_1,s_2,\ldots,s_6) \in [0,1]^6$, for $i=1,2,\ldots,6$, and corresponding offspring PGFs for the initial generation of the forward branching process.  In the initial generation there are two types of individual, $U$ and $V$, depending on whether the initial infective in the epidemic process is unvaccinated or vaccinated, respectively.  For $A \in \{ U,V \}$, let $\bs{C}_A = (C_{A1}, C_{A2}, \ldots, C_{A6})$ denote the offspring random variable for a type-$A$ initial individual in the forward branching process and let $f_{\bs{C}_A} (\bs{s}) = E[\bs{s}^{\bs{C}_A}]$ denote the PGF of $\bs{C}_A$.

Recall that secondary individuals in a household are vaccinated independently, each with probability $p_V$.  Thus, conditioning first on household size and then on the number of secondary members that are vaccinated yields, for $i=1,2,\ldots,6$,
\[
f_{\tilde{\bs{C}}_i} (\bs{s}) = \sum_{n=1}^{\infty} \tilde{\rho}_n \sum_{v_s=0}^{n-1} \binom{n-1}{v_s} p_V^{v_s} (1-p_V)^{n-1-v_s}
f_{ \tilde{\bs{C}}_i^{(n,v_s)}} (\bs{s}),
\]
where $\tilde{\bs{C}}_i^{(n,v_s)}$ is the offspring random variable for a type-$i$ non-initial individual, given that the corresponding household is of size $n$ and has $v_s$ secondary members vaccinated.  Similarly, and in an obvious notation, for $A \in \{ U,V \}$,
\[
f_{\bs{C}_A} (\bs{s}) = \sum_{n=1}^{\infty} \tilde{\rho}_n \sum_{v_s =0}^{n-1} \binom{n-1}{v_s} p_V^{v_s} (1-p_V)^{n-1-v_s}
f_{\bs{C}_A^{(n,v_s)}} (\bs{s}).
\]

With an all-or-nothing vaccine action, we may condition also on the number of vaccinations of secondary members that are unsuccessful to obtain, for $i=1,2,\ldots,6$,
\[
f_{\tilde{\bs{C}}_i^{(n,v_s)}}(\bs{s})=\sum_{u_s=0}^{v_s} \binom{v_s}{u_s} (1-\varepsilon)^{u_s} \varepsilon^{v_s-u_s} f_{\hat{\bs{C}}_i^{(n-v_s+u_s,u_s)}} (\bs{s}),
\]
where $\hat{\bs{C}}_i^{(n',v')}$ is the offspring random variable for a type-$i$ non-initial individual, given that the corresponding household contains $n'-1$ other susceptibles, of which $v'$ are unsuccessfully vaccinated.  (Note that, unlike with households based vaccination, the vaccine status of these susceptibles is important as it affects their degree distributions.)  Similarly, and in an obvious notation, for $A \in \{ U,V \}$,
\[
f_{\bs{C}_A^{(n,v_s)}}(\bs{s}) = \sum_{u_s=0}^{v_s} \binom{v_s}{u_s} (1-\varepsilon)^{u_s} \varepsilon^{v_s-u_s} f_{\hat{\bs{C}}_A^{(n-v_s+u_s,u_s)}} (\bs{s}).
\]

Calculation of $f_{\tilde{\bs{C}}_i^{(n,v_s)}} (\bs{s})$ and $f_{\bs{C}_A^{(n,v_s)}} (\bs{s})$ when the vaccine action is non-random and of $f_{\hat{\bs{C}}_i^{(n',v')}}(\bs{s})$ and $f_{\hat{\bs{C}}_A^{(n',v')}} (\bs{s})$ when the vaccine action is all-or-nothing is described in Appendix~\ref{AppAcqVac}. As with previous calculations, these simplify appreciably if the infectious period is constant.

In either case we approximate the probability of a major outbreak, with a single initial infective being of type $A\in\{U,V\}$, as $\pmaj^{(A)}=1-f_{\bs{C}_A}(\bs{\sigma})$, where $\bs{\sigma}=(\sigma_i,\,i=1,2,\dots,6)$ is the unique solution of $f_{\tilde{\bs{C}}_i}(\bs{\sigma})=\sigma_i$ ($i=1,2,\dots,6$) in $[0,1)^6$. (This assumes of course that $R_v>1$. If $R_v\leq1$ then $\pmaj^{(V)}=\pmaj^{(U)}=0$.) The probability of a major outbreak with an initial infective chosen uniformly from the whole population is therefore
\begin{equation*}
\pmaj = \begin{cases}
p_V \pmaj^V + (1-p_V)\pmaj^U, & \mbox{non-random vaccine,}\\
p_V (1-\epsilon) \pmaj^V + (1-p_V)\pmaj^U, & \mbox{all-or-nothing vaccine.}
\end{cases}
\end{equation*}
(With an all-or-nothing vaccine, if the initial infective is vaccinated then a major outbreak can occur only if that vaccination is unsuccessful, which occurs with probability $1-\epsilon$.)

\subsection{Final outcome of major outbreak}
\label{subsecacqfinal}

We approximate the susceptibility set of a given individual, $i^{\ast}$ say, by a households-based (backward) multitype branching process, where, as in Section~\ref{subsechousevacnr}, the type of a household is given by the type of the primary member of the corresponding local susceptibility set.  In the initial generation there are two types, $V$ and $U$, depending on whether or not $i^{\ast}$ is vaccinated.  In subsequent generations, there are 6 types, numbered 1--6 as in Section~\ref{subsecacqmod}, where now $N$ means that the primary case was named by the individual that it contacts globally to join $i^{\ast}$'s susceptibility set.  Similar to before, secondary members of a local susceptibility set need only to be typed $V$ or $U$.  Let
$\bs{B}_U = (B_{U1}, B_{U2},\ldots,B_{U6})$ and $\bs{B}_V=(B_{V1},B_{V2},\ldots,B_{V6})$ denote the offspring random variables for the initial generation of this branching process, and let $\tilde{\bs{B}}_i = (\tilde{B}_{i1}, \tilde{B}_{i2}, \ldots, \tilde{B}_{i6})$ ($i=1,2,\ldots,6$) be the offspring random variables for all subsequent generations.

We determine first the PGF $f_{\tilde{\bs{B}}_i} (\bs{s})$, where $\bs{s}=(s_1,s_2,\ldots,s_6) \in [0,1]^6$.  Let
$\bs{M}_i = (M_{iU}, M_{iV})$, where $M_{iU}$ and $M_{iV}$ are respectively the number of unvaccinated and vaccinated secondary individuals in the local susceptibility set of a typical type-$i$ primary individual.  Then $\tilde{\bs{B}}_i$ admits the decomposition
\begin{equation}
\tilde{\bs{B}}_i = \hat{\bs{B}}_i^P + \sum_{k=1}^{M_{iU}} \hat{\bs{B}}_U^S (k) + \sum_{l=1}^{M_{iV}} \hat{\bs{B}}_V^S (l), \label{Btildei}
\end{equation}
where $\hat{\bs{B}}_i^P$, $\hat{\bs{B}}_U^S (k)$ and $\hat{\bs{B}}_V^S (l)$ are the respective contributions to $\tilde{\bs{B}}_i$ from the primary individual, the $k$th unvaccinated secondary individual and the $l$th vaccinated secondary individual in the local susceptibility set.  The random variables being summed on the right hand side of (\ref{Btildei}) are mutually independent and independent of $\bs{M}_i$, so, in an obvious notation,
\begin{equation}
f_{\tilde{\bs{B}}_i} (\bs{s}) = f_{\bs{B}_i^P}(\bs{s}) f_{\bs{M}_i} (f_{\hat{\bs{B}}_U^S} (\bs{s}), f_{\hat{\bs{B}}_V^S} (\bs{s})).\label{fBitilde}
\end{equation}

To derive the PGF $f_{\bs{M}_i}$, note that $f_{\bs{M}_1}=f_{\bs{M}_2} = f_{\bs{M}_4} = f_{\bs{M}_5}$ ($=f_{\bs{M}_V}$ say) and
$f_{\bs{M}_3} = f_{\bs{M}_6}$ ($=f_{\bs{M}_U}$ say).  Further, conditioning on the size of the household and noting that secondary individuals in the household are vaccinated independently, each with probability $p_V$, yields
\[
f_{\bs{M}_A} (\bs{s}) = \sum_{n=1}^{\infty} \tilde{\rho}_n \sum_{v=0}^{n-1} \binom{n-1}{v} p_V^v (1-p_V)^{n-1-v} f_{\bs{M}_A^{(n,v+\delta_{A,V})}}
(\bs{s}) \quad (\bs{s}\in[0,1]^2)
\]
for $A \in \{ U,V \}$, where $\bs{M}_A^{(n,v)}$ is as in Section~\ref{subsechousevacnr}.  Calculation of $f_{\bs{M}_A^{(n,v)}} (\bs{s})$ for the two models of vaccine action is described in Appendix~\ref{AppSusset}.

We determine next the PGFs $f_{\hat{\bs{B}}_i^P}$ ($i=1,2,\ldots,6$).  For $i=1,2,\ldots,6$, let $\tilde{D} (i)$ denote the degree of a typical type-$i$ primary individual, who thus has $\tilde{D}(i)-1$ global neighbours in the construction of the backward process, and let
$\tilde{\bs{X}}_i = (\tilde{X}_{i1}, \tilde{X}_{i2} , \ldots, \tilde{X}_{i6})$, where $\tilde{X}_{ij}$ is the number of those
$\tilde{D}(i)-1$ global neighbours that are of type $j$, so $\sum_{j=1}^6 \tilde{X}_{ij} = \tilde{D}(i)-1$. To determine the PGF $f_{\hat{\bs{B}}_i^P}$ we first condition on $\tilde{\bs{X}}_i$ and then consider how many of the $\tilde{\bs{X}}_i$ individuals of the various types actually join the susceptibility set.  For $i=1,3,4$ and $6$, the types of these $\tilde{D}(i)-1$ global neighbours are chosen independently, according to $(\hat{p}_{ij},\, j=1,2,\ldots,6)$, so
\[
f_{\tilde{\bs{X}}_i} (\bs{s}) = f_{\tilde{D}(i)-1} (\hat{g}_i (\bs{s})) \quad (i=1,3,4,6),
\]
where $\hat{g}_i (\bs{s}) = \sum_{j=1}^6 \hat{p}_{ij} s_j$.  Further $\tilde{D}(1) \stackrel{D}{=} \tilde{D}(4) \sim \tilde{D}$ and
$\tilde{D}(3) \stackrel{D}{=} \tilde{D}(6) \sim \tilde{D}_U$ (see (\ref{pdtildeu})), where $\stackrel{D}{=}$ denotes equality in distribution. 

Now consider a typical type-$2$ primary individual and let $\tilde{N}_S$ and $\tilde{N}_{S^c}$ be the number of its $\tilde{D}(2)-1$ global neighbours
in the construction of the backward process that are sampled and unsampled, respectively, so $\tilde{N}_S + \tilde{N}_{S^c}=\tilde{D}(2)-1$.  Now each of these $\tilde{D}(2)-1$ neighbours independently is named with probability $p_N$ and, if it is not named, it is vaccinated with probability $\tilde{p}_V$, so
\[
f_{\tilde{\bs{X}}_2} (\bs{s}) = E_{\tilde{N}_S, \tilde{N}_{S^c}} \left[(\hat{g}_{21} (\bs{s}))^{\tilde{N}_S} (\hat{g}_{22} (\bs{s}))^{\tilde{N}_{S^c}}\right],
\]
where $\hat{g}_{21} (\bs{s}) = p_N s_1 + (1-p_N) \tilde{p}_V s_2 + (1-p_N )(1-\tilde{p}_V) s_3$ and
$\hat{g}_{22} (\bs{s}) = p_N s_4 + (1-p_N) \tilde{p}_V s_5 + (1-p_N) (1-\tilde{p}_V)s_6$.  Further, $\tilde{D}(2) \sim \tilde{D}_V$, hence using (\ref{pdtildev}) and (\ref{ptildensdv}), we obtain
\begin{equation}
f_{\tilde{\bs{X}}_2} (\bs{s}) = \tilde{p}_V^{-1} \left[ f_{\tilde{D}-1} \left(\hat{g}_2 (\bs{s}, 0) \right) - f_{\tilde{D}-1} \left(\hat{g}_2 (\bs{s}, p_N )\right)\right],
\label{fX2tilde}
\end{equation}
where $\hat{g}_2 (\bs{s},x)=p_S (1-x) \hat{g}_{21} (\bs{s}) + (1-p_S) \hat{g}_{22} (\bs{s})$.  A similar argument shows that
\begin{equation}
f_{\tilde{\bs{X}}_5} (\bs{s}) = \tilde{p}_V^{-1} \left[ f_{\tilde{D}-1} \left(\hat{g}_5 (\bs{s},0) \right) - f_{\tilde{D}-1} \left( \hat{g}_5 (\bs{s},p_N) \right)\right],
\label{fX5tilde}
\end{equation}
where $\hat{g}_5 (\bs{s},x) = p_S (1-x) \hat{g}_{51} (\bs{s}) + (1-p_S) \hat{g}_{52} (\bs{s})$, with
$\hat{g}_{51} (\bs{s}) = \tilde{p}_V s_2 + (1-\tilde{p}_V) s_3$ and $\hat{g}_{52} (\bs{s}) = \tilde{p}_V s_5 + (1-\tilde{p}_V)s_6$.

Consider a typical type-$i$ primary individual.  Each of its $\tilde{D}(i)-1$ global neighbours in the construction of the backward process enters the susceptibility set independently and with probability $p_{ij}^B$, where $j$ is the type of the global neighbour. Here $p_{ij}^B$ is the probability that a type-$j$ neighbour of a type-$i$ individual joins the susceptibility set. It follows that
\begin{equation*}
p_{ij}^B = \begin{cases}
p_{ji}^G & \mbox{non-random vaccine,} \\
p_{ij}^G & \mbox{all-or-nothing vaccine.}
\end{cases}
\end{equation*}
(The different formulae arise from the fact that in the all-or-nothing case, in addition to requiring a contact from the neighbour to the type-$i$ individual of interest, a vaccinated neighbour is able to join the susceptibility set only if its vaccination fails.) Hence, for $i=1,2,\ldots,6$,
\[
f_{\hat{\bs{B}}_i^P} (\bs{s}) = f_{\tilde{\bs{X}}_i} (\bs{h}_i^B (\bs{s})),
\]
where $\bs{h}_i^B (\bs{s}) = (h_{i1}^B (s_1), h_{i2}^B (s_2), \ldots, h_{i6}^B (s_6))$, with $h_{ij}^B (s) = 1-p^G_{ij} + sp^G_{ij}$ ($j=1,2,\ldots,6$).

We determine now the PGFs $f_{\hat{\bs{B}}_U^S}$ and $f_{\hat{\bs{B}}_V^S}$.  In an obvious notation,
\[
f_{\hat{\bs{B}}_A^S} (\bs{s}) = f_{\bs{X}_A} (\bs{h}_A^B (\bs{s})) \quad (A \in \{U,V\}),
\]
where $\bs{h}_U^B = \bs{h}_3^B$ and $\bs{h}_V^B = \bs{h}_1^B$.  A typical type-$U$ secondary individual, $j_U^{\ast}$ say, has degree distributed according to $D_U$ (recall (\ref{pDUd})) and, since $j_U^{\ast}$ enters the susceptibility set process through a local susceptibility set, all of $j_U^{\ast}$'s global neighbours are available to join the susceptibility set.  Further, $j_U^{\ast}$ is sampled with probability $p_S$, in which case, apart from its degree, $j_U^{\ast}$ behaves similarly to a type-3 primary individual, otherwise $j_U^{\ast}$ is not sampled and behaves similarly to a type-6 primary individual.  Thus,
\[
f_{\bs{X}_U} (\bs{s}) = p_S f_{D_U} (\hat{g}_3 (\bs{s})) + (1-p_S) f_{D_U} (\hat{g}_6 (\bs{s})).
\]
A typical type-$V$ secondary individual, $j_V^{\ast}$ say, is also sampled with probability $p_S$, and a similar argument, using appropriate modifications of (\ref{fX2tilde}) and (\ref{fX5tilde}) as $j_V^{\ast}$ has degree distributed according to $D_V$, yields
\begin{align*}
f_{\bs{X}_V} (\bs{s}) &= p_V^{-1} \left\{ p_S \left[f_D \left(\hat{g}_2 (\bs{s},0)\right) - f_D \left(\hat{g}_2 (\bs{s},p_N)\right)\right]\right.\\
&\qquad \qquad
+\left. (1-p_S)\left[f_D\left(\hat{g}_5 (\bs{s},0)\right) - f_D\left(\hat{g}_5 (\bs{s},p_N)\right)\right]\right\}.
\end{align*}

Finally, we determine the offspring PGFs for the initial generation, $f_{\bs{B}_U}$ and $f_{\bs{B}_V}$.  Observe that, for
$A \in \{ U,V \}$, in the initial generation, a primary individual of type $A$ behaves according to the same probability law as a typical secondary type-$A$ individual (in any generation), so (\ref{fBitilde}) becomes
\[
f_{\bs{B}_A} (\bs{s}) = f_{\hat{\bs{B}}_A^S} (\bs{s}) f_{\bs{M}_A} \left( f_{\hat{\bs{B}}_U^S} (\bs{s}), f_{\hat{\bs{B}}_V^S} (\bs{s}) \right) \quad
(A \in \{ U,V \})
\]
and $f_{\bs{B}_A} (\bs{s})$ can be evaluated using results given above.

Suppose that $0 < p_V < 1$, $p_G (U,V) p_G (V,U) \neq 0$ and $R_v > 1$.  Let $\tilde{\bs{\xi}} = ( \tilde{\xi}_1, \tilde{\xi}_2, \ldots, \tilde{\xi}_6 )$ be the unique solution in $[0,1)^6$ of the equations $f_{\tilde{\bs{B}}_i} (\bs{s}) = s_i$ ($i=1,2,\ldots,6$).  Then the proportions of unvaccinated and vaccinated individuals that are infected by a major epidemic are given by $z^U = 1-f_{\bs{B}_U} (\tilde{\bs{\xi}})$ and $z^V = 1 - f_{\bs{B}_V} (\tilde{\bs{\xi}})$, respectively. The overall proportion of the population infected by a major outbreak is therefore
\begin{equation*}
z = \begin{cases}
p_V z^V + (1-p_V)z^U, & \mbox{non-random vaccine,}\\
p_V (1-\epsilon) z^V + (1-p_V)z^U, & \mbox{all-or-nothing vaccine.}
\end{cases}
\end{equation*}

\section{Model behaviour/exploration}
\label{secbehaviour}

In this section we explore the behaviour of the model and its dependence on some key parameters. The main focus is on numerical comparison of the various vaccine allocation regimes for a variety of household size and network degree distributions. First, however, we comment briefly on the relationship between outcomes in simulated finite populations and our asymptotic analytical results, and also give some discussion of the choice of the parameters $p_S$ and $p_N$ in acquaintance vaccination.% Some analytical progress is possible when we look at the effect of the balance of $p_S$ and $p_N$ while fixing coverage if the vaccine is perfect. Numerically, we explore how well the properties of the epidemic processes we consider are approximated by our asymptotic analytical results and also briefly investigate how the relative effectiveness of differernt vaccine allocation regimes depends on model parameters.

\subsection{Outcomes in finite populations}
One can compare the analytical, asymptotic, quantities of interest ($\pmaj$ and $z$) to simulation-based estimates of the corresponding quantities on finite populations. As in \citeA[Section~5]{BalSirTra2009}, we find that the agreement becomes reasonable with the number of households $m$ in the low hundreds and very close indeed for $m$ over about 1000, though the convergence is a little slower for heavier tailed degree distributions. %\footnote{Also comment that the coverage converges in this way. Since we specify $p_S$ and $p_N$ we find that in small populations well-connected individuals tend to be named and on average the proportion of the population that are actually vaccinated is slightly less than $c$.}

\subsection{Acquaintance vaccination dependence on $p_S$ and $p_N$}
\label{subsecacqpSpN}
The vaccine coverage $c=p_V=1-f_D(1-p_Sp_N)$ in our acquaintance vaccination model is determined by the product $p_Sp_N$ (see equation~(\ref{pV})). Here we investigate the dependence of the model outcomes ($R_v$, $\pmaj$ and $z$) on the precise values of $p_N$ and $p_S$, for a fixed coverage. First we consider a perfect vaccine, in which case analytical progress is possible. The argument below follows much the same lines as the corresponding argument for the standard network model~\cite{BalSir2013}. 

Since the vaccine is perfect, we have only households of types 3 and 6 (i.e.\ $(U,S)$ and $(U,S^c)$); and since these two types are sampled independently with the same probability, both the backward and forward processes reduce to single-type. The mean number of forward neighbours of an infected individual is $\mu_{\tilde{D}_U-1} + \mu_S(U,U)\mu_{D_U}$, where $\mu_S(U,U)$ is (as in earlier sections) the mean number of unvaccinated secondary cases in a single-household epidemic initiated by an unvaccinated primary case. Each infected individual is sampled with probability $p_S^U$, so fails to name each forward neighbour with probability $p_S^U(1-p_N) + 1-p_S^U$. Then each such unnamed individual avoids vaccination (by other neighbours in the network) with probability $1-\tilde{p}_V$. Thus
\begin{align*}
R_v & = (\mu_{\tilde{D}_U-1} + \mu_{D_U}\mu_S(U,U)) (p_S^U(1-p_N) + 1-p_S^U) (1-\tilde{p}_V) p_G.
\end{align*}
Now writing $p'=p_Sp_N$ for the part of $(p_S,p_N)$ which specifies the coverage, and using~\eqref{pVtilde},~\eqref{muDUtilde1} and~\eqref{muDU}, this can be written as
\begin{align*}
R_v & = f_{\tilde{D}-1}(1-p') p_G \left( \frac{f'_{\tilde{D}-1}(1-p')}{f_{\tilde{D}-1}(1-p')} + \mu_S(U,U)\frac{f'_D(1-p')}{f_D(1-p')}\right) (1-2p'+p'p_N) \\
 & = \eta(p') (1-2p'+p'p_N),
\end{align*} 
where the function $\eta$ depends on $p_S$ and $p_N$ only through their product $p'$. It is immediate that $R_v$ is increasing in $p_N$ for fixed $p'=p_Sp_N$. Thus we have that the best effect is achieved by taking $(p_S,p_N)=(1,p')$ and the worst with $(p_S,p_N)=(p',1)$; with $R_v^b=\eta(p')(1-p')^2$ and $R_v^w=\eta(p')(1-p')$, i.e.\ $R_v^b = (1-p')R_v^w$. Precisely as in the network-only model without households, we note that this difference is generally quite small and is at its largest when $p'$ is large, so there is high coverage and the epidemic is likely to be subcritical in any event.

When the vaccine action is not perfect we cannot make analytical progress in this direction, but extensive numerical calculations suggest that $R_v$, $\pmaj$ and $z$ are usually monotonic in $p_N$ when $p_Sp_N$ is fixed and that in any case the difference in outcomes between the best and worst choices are very small. We do not explicitly demonstrate this here, but we note that this is the same as found in \citeA{BalSir2013} for the model without households and can be seen to some extent in all of the plots in Section~\ref{secnumerics}, where the two acquaintance vaccination plots are barely distinguishable in most cases.

In the further numerical studies below we use the terms `best' and `worst' to refer to the acquaintance vaccination schemes with $(p_S,p_N)=(1,p')$ and $(p_S,p_N)=(p',1)$ even though the names are not necessarily correct. As noted above, the precise choice of $p_S$ and $p_N$ seems to have only a very weak influence on the final outcome of the epidemic model and these two cases seem to give bounds for the quantities of interest.

\subsection{Effect of different vaccine allocation regimes}
\label{secnumerics}
We now explore the relative effectiveness of the different vaccine allocation regimes discussed in this paper, across a small but representative variety of household size and degree distributions. In the following figures we plot an outcome ($R_v$ or $z$) against the vaccine coverage $c$ for several vaccine allocation strategies; with all other parameters held fixed. Using different plots within the same figure we vary some of these other parameters. The vaccine allocation methods we consider are (i) uniformly chosen individuals (Ind UAR), (ii) uniformly chosen households (HH UAR), (iii \& iv) best and worst household-based allocation (HH Best and HH Worst), (v \& vi) `best' and `worst' acquaintance vaccination (Acq Best and Acq Worst); see the end of Section~\ref{subsechousevacintro} for descriptions of the household-based vaccination schemes.

In this numerical section we use household size distributions $\rhoUK=(31,32,16,14,5,2)/100$ and $\rhoPak=(9,45,77,118,141,146,123,104,237)/1000$, which have respective means 2.4 and 6.8, as realistic household size distributions from the UK~\cite[Table 3.1]{ONSGLS2011} and Pakistan~\cite[Table~2.9]{PakDHS2013}, respectively. The degree distributions we use are the standard Poisson distribution and a power law distribution with exponential cutoff (as in, for example,~\citeA{Amaraletal2000}). For this latter distribution we use the notation $D\sim\mbox{PowC}(a,\kappa)$ to denote that $p_k = k^{-a} \re^{-k/\kappa}$ ($k=1,2,\dots$), i.e.\ a power law with index $a$ and exponential cutoff at about $\kappa$. In particular, we use the degree distribution $D\sim\mbox{PowC}(2,120)$, which has mean $\mu_D\approx3.001$ and $\sigma^2_D\approx66$. We also use the notation $\mbox{Gam}(k,r)$ to denote a Gamma distribution with shape parameter $k$ and scale parameter $r$ (and thus mean $kr$ and variance $kr^2$). %Throughout this section we use the notation $c$ instead of $p_V$, since the interpretation as the coverage is the emphasis here.

\begin{figure}
\begin{center}
\includegraphics[width=\hfigwidth]{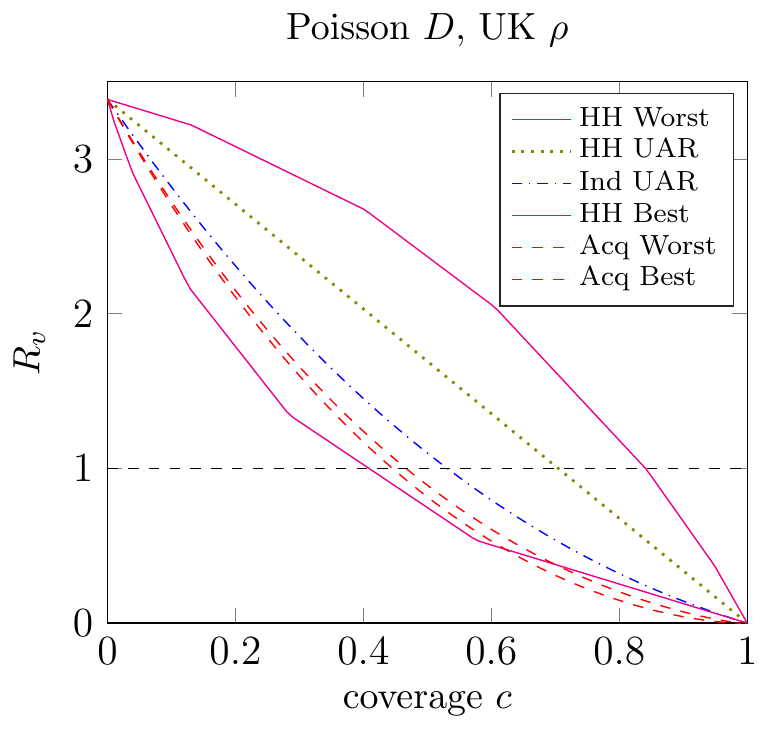}
\includegraphics[width=\hfigwidth]{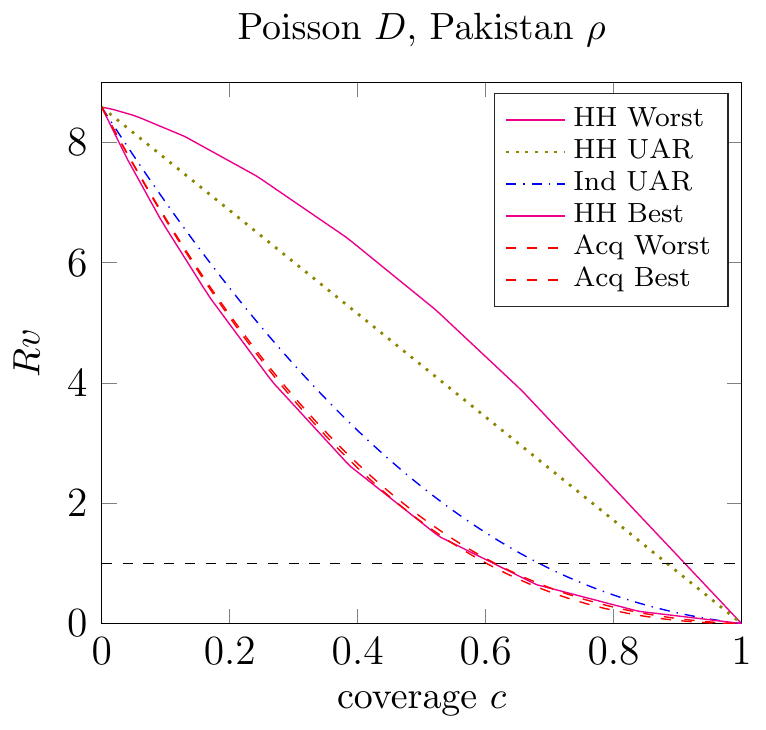}
\\[0.3cm]
\includegraphics[width=\hfigwidth]{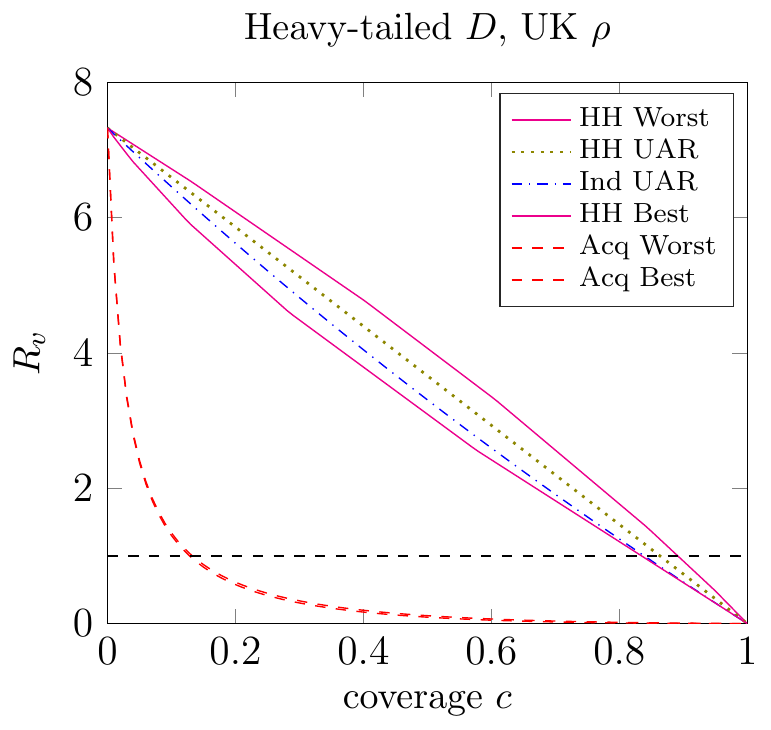}
\includegraphics[width=\hfigwidth]{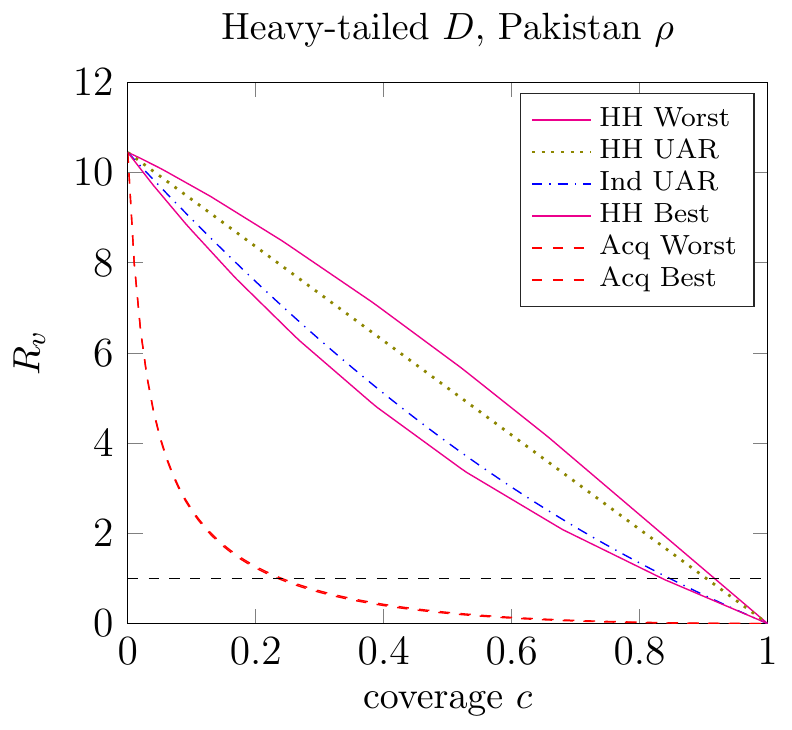}
\end{center}
\caption{Plots of $R_v$ versus coverage ($c$) for various kinds of allocation of a perfect vaccine. We use two different household size distributions: upper plots $\rho=\rhoUK$, lower plots $\rho=\rhoPak$; and two different degree distributions: left plots $D\sim\mbox{Poi}(5)$, right plots $D\sim\mbox{PowC}(2,120)$. Other parameters are $I\sim\mbox{Gam}(5,1/5)$, $\lambda_L=1$, $\lambda_G=0.3$.}
\label{fig:RvVcPerfect}
\end{figure}

First we look at the relative effectiveness of the various allocation schemes, and how this changes with the household size distribution and the network degree distribution. Figure~\ref{fig:RvVcPerfect} shows plots of the post-vaccination threshold parameter $R_v$ as a function of the vaccine coverage $c$ for the 6 vaccine allocation schemes with a perfect vaccine. The different plots use different combinations of household size and network degree distribution, with all other parameters kept fixed (full details are in the figure caption). Firstly we note that the patterns of the household-based allocation schemes are consistent with those of~\citeA[Section~4]{BalLyn2006} for the standard households model. We see that when the network degree distribution is not very variable (Poisson) good household-based schemes perform similarly to acquaintance vaccination. On the other hand, when the network degree distribution is much more variable (cut-off power law) we see that acquaintance vaccination significantly outperforms the household-based schemes; though to a slightly lesser extent when the household size distribution is also more variable.

\begin{figure}
\begin{center}
\includegraphics[width=\hfigwidth]{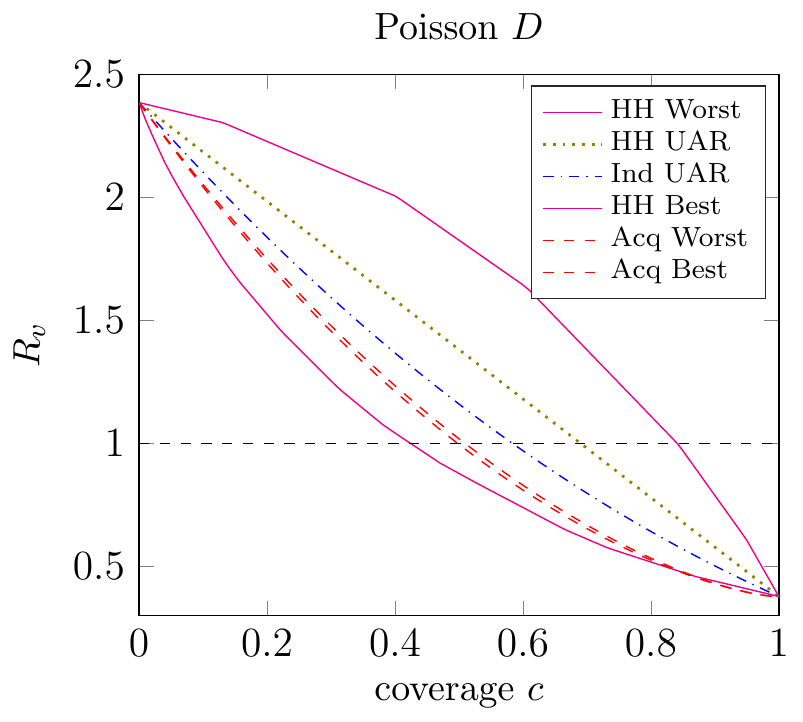}
\includegraphics[width=\hfigwidth]{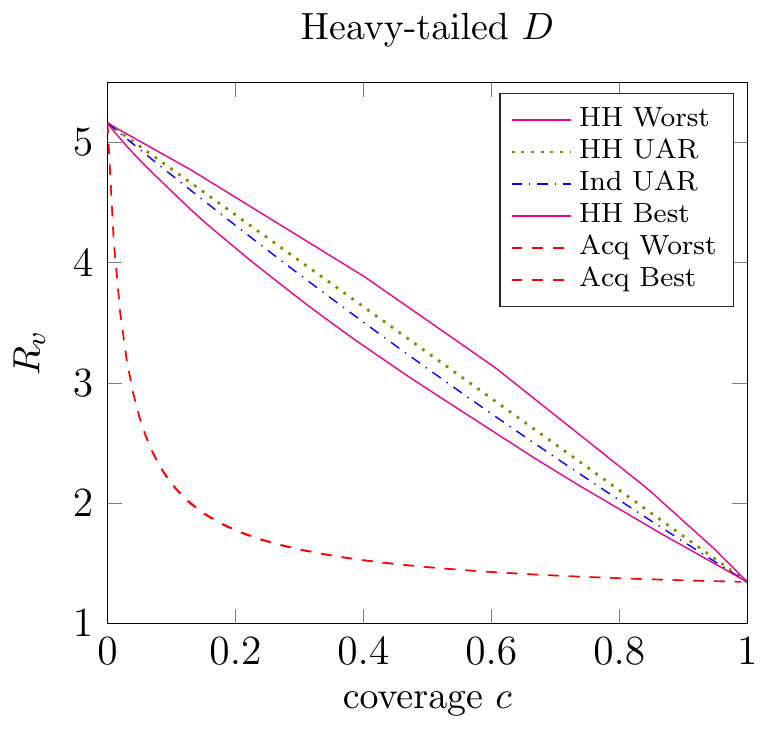}
\\[0.3cm]
\includegraphics[width=\hfigwidth]{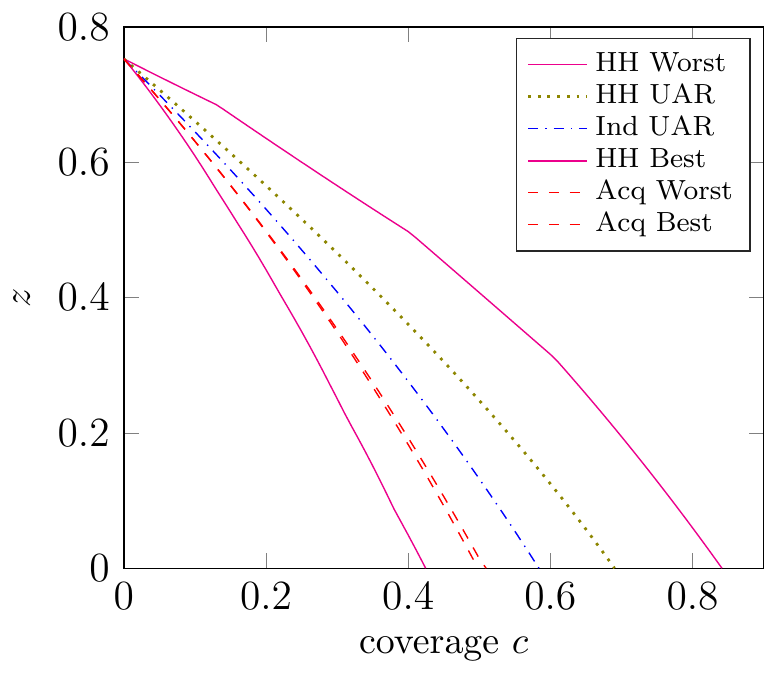}
\includegraphics[width=\hfigwidth]{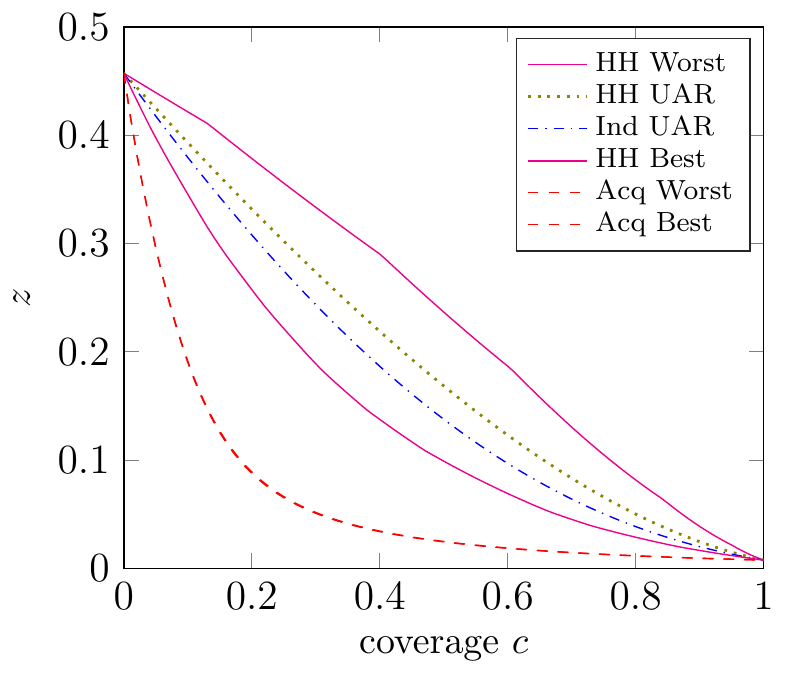}
\end{center}
\caption{Plots of $R_v$ and $z$ versus coverage ($c$) for various kinds of allocation of an all-or-nothing vaccine with success probability $\epsilon=0.7$. We use two different degree distributions: left plots $D\sim\mbox{Poi}(5)$, right plots $D\sim\mbox{PowC}(2,120)$. Other parameters are $\rho=\rhoUK$, $I\sim\mbox{Gam}(5,1/5)$, $\lambda_L=1$, $\lambda_G=0.2$.}
\label{fig:RvandzVcAoN}
\end{figure}

\begin{figure}
\begin{center}
\includegraphics[width=\hfigwidth]{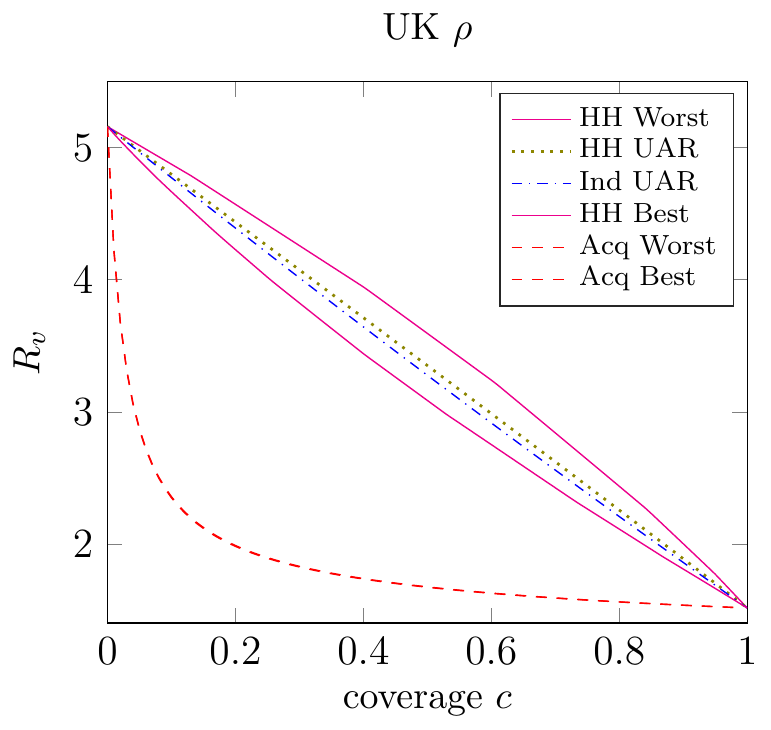}
\includegraphics[width=\hfigwidth]{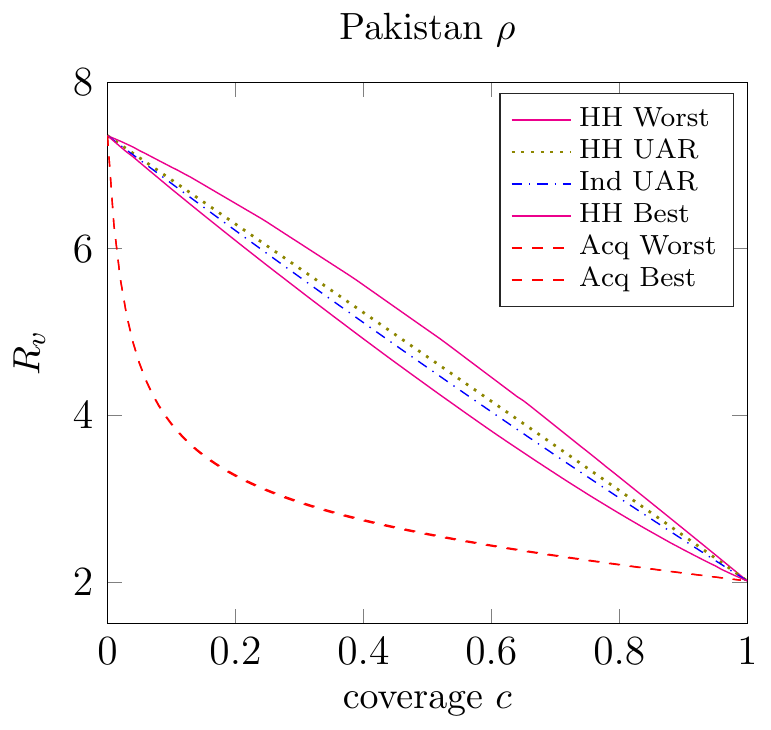}
\\[0.3cm]
\includegraphics[width=\hfigwidth]{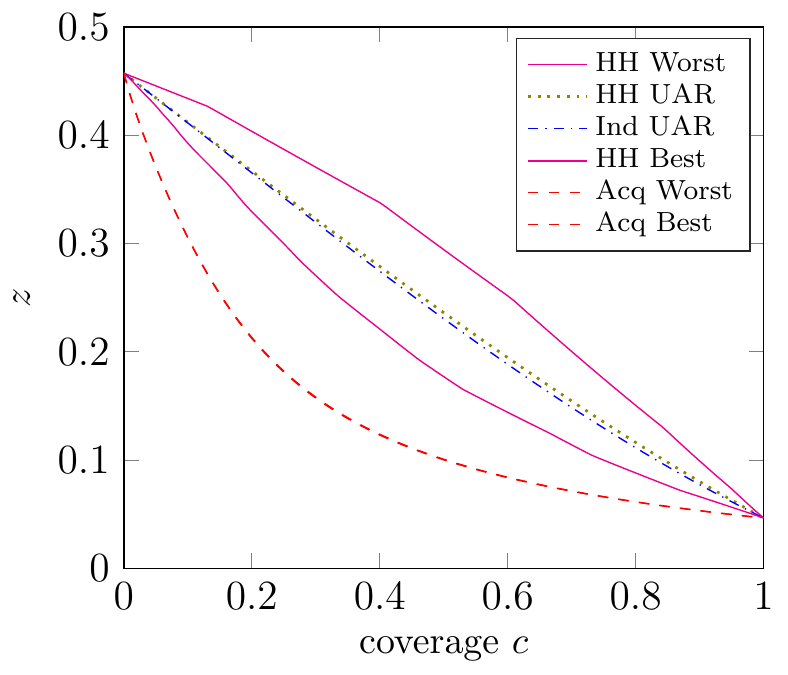}
\includegraphics[width=\hfigwidth]{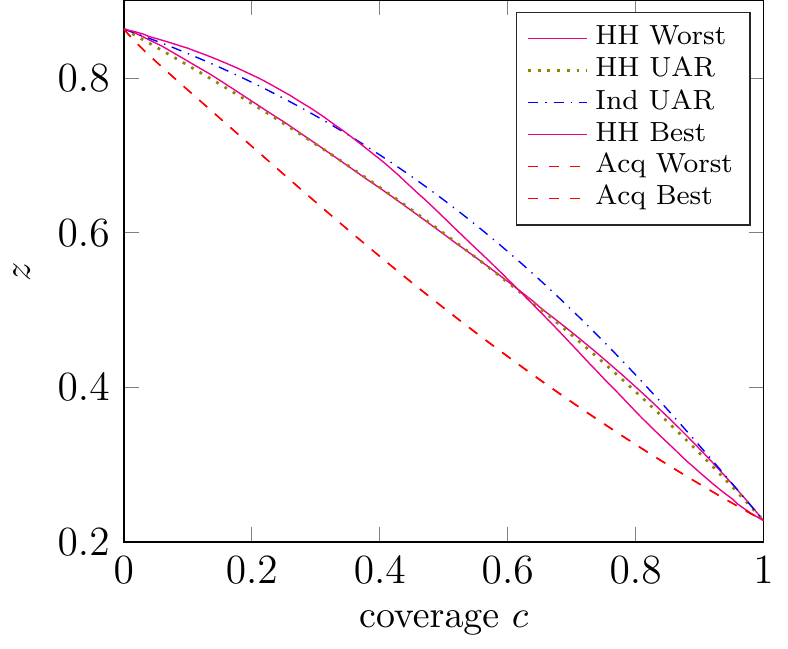}
\end{center}
\caption{Plots of $R_v$ and $z$ versus coverage ($c$) for various kinds of allocation of a non-random vaccine with relative susceptibility and infectivity $a=0.5$ and $b=0.6$ respectively. We use two different household size distributions: left plots $\rho=\rhoUK$, right plots $\rho=\rhoPak$. Other parameters are $D\sim\mbox{PowC}(2,120)$, $I\sim\mbox{Gam}(5,1/5)$, $\lambda_L=1$, $\lambda_G=0.2$.}
\label{fig:RvandzVcNR}
\end{figure}

Figures~\ref{fig:RvandzVcAoN} and~\ref{fig:RvandzVcNR} are similar to Figure~\ref{fig:RvVcPerfect}, but we consider imperfect vaccine action models (with the same \emph{efficacy} $1-E[AB]=0.7$) and we plot both the post-vaccination threshold parameter $R_v$ and expected final size of a large outbreak $z$. In Figure~\ref{fig:RvandzVcAoN} we use an all-or-nothing vaccine action model with success probability $\epsilon=0.7$ and consider two network degree distributions, with fixed (less variable) household size distribution. In Figure~\ref{fig:RvandzVcNR} we use a non-random vaccine action model with relative susceptibility and infectivity $a=0.5$ and $b=0.6$, respectively, and vary the household size distibution, with a fixed (more variable) network degree distribution. Note that for this choice of $a$ and $b$, determining the `best' and `worst' household-based vaccine allocation schemes are not linear programming problems (see Section~\ref{subsecopt}). In our numerical routine we use the MATLAB constrained optimisation solver {\tt fmincon}.

We see in the upper plots (of $R_v$) in Figures~\ref{fig:RvandzVcAoN} and~\ref{fig:RvandzVcNR} broadly similar patterns to those in Figure~\ref{fig:RvVcPerfect}. The lower plots in Figure~\ref{fig:RvandzVcAoN} show the ordering of $z$ for the different allocation regimes being the same as the ordering of $R_v$. These two lower plots look qualitatively quite different, but it is important to note that in one the vaccine can and in the other the vaccine cannot bring the epidemic below threshold. These plots suggest that, as might be expected, when the network degree distribution is not so variable (e.g.\ the Poisson case) then vaccine allocation should be focussed on households-based methods, whilst when the network degree distribution is more variable (e.g.\ the cutoff power law) then targeting vaccination effort based on the network might give better results. Precisely which method is preferable will of course depend heavily on the other parameters of the model, but we have demonstrated that either allocation method, household-based or acquaintance-based, can be preferable to the other.

The lower plots of Figure~\ref{fig:RvandzVcNR} follow similar patterns in that we are considering the case of a quite variable network degree distribution so acquaintance vaccination outperforms the household-based methods. There are however some unexpected patterns in the lower-right plot of Figure~\ref{fig:RvandzVcNR}, in that the ordering of the various household-based allocation regimes are not the same as in the corresponding plot of $R_v$ immediately above it. In particular, for lower coverages the `worst' households based allocation outperforms the `best'! This demonstrates that when the epidemic is well above threshold, optimising vaccine allocation based on $R_*$ does not necessarily result in an expected final size that is as low as possible; cf.\ \citeA{KeeRos2015}, who observe a similar phenomenon in the standard households model with a perfect vaccine. The threshold parameter $R_v$ measures household-to-household transmission, but does not directly take into account the size of the within-household outbreaks. This could perhaps be resolved by optimising an individual-based threshold parameter instead (see \citeA{PelBalTra2012,BalPelTra2016}), but the optimisation problem would be more difficult than the one we have considered. We also note that this phenomenon appears to arise only when the epidemic is well above criticality.

Lastly we note here that the behaviour of $\pmaj$ is broadly similar to that of $z$. We do not present any plots here, but they have shapes and patterns similar to those in the $z$ plots that are shown, including the unexpected ordering observed in the lower-right plot of Figure~\ref{fig:RvandzVcNR}.

%, though the scales are slightly different. The ordering of the performance of the different allocation strategies appears to be the same for these two measures of severity; i.e.\ the coverages at which two vaccination strategies are equally effective in the lower-right plot of Figure~\ref{fig:RvandzVcNR} are precisely the same if one considers $\pmaj$ instead of $z$ as the measure of performance of the vaccination.

\section{Concluding remarks}
\label{secconclusion}

In this paper we have analysed vaccine allocation strategies in stochasitc SIR epidemic models upon populations with household and random network (of the configuration model type) structure. By exploiting branching process approximations we derive asymptotic results describing the threshold behaviour of this epidemic model when there are few initial infectives and the final outcome in the event of a major outbreak. Particular attention has been paid to the analysis of acquaintance vaccination, which aims to target vaccination at individuals who are more highly connected in the network. We find that acquaintance vaccination potentially offers substantial benefits over other (households-based) vaccine allocation regimes.

Whilst we have shown that acquaintance vaccination is potentially useful, it is clearly not feasible to implement in practice in human populations, so investigation of a more ethically acceptable allocation regime that preserves the targeting of well-connected individuals is a clear direction for future work. It is also likely that the effectiveness of acquaintance vaccination will depend on the amount of clustering in the network in an interesting way; we have touched on this through the use of different household size distributions, but clearly there is scope for considerably more work in this direction. Further issues that warrant more investigation include comparison to optimal configuration model based vaccine allocation, assuming knowledge of the degree of every individual in the network (see~\citeA[Section 5 and Appendix B]{BalSir2012}), the determination of good/optimal vaccination strategies based on household and network information and extending acquaintance vaccination to include the possiblity of naming individuals who are in the same household.

A particularly striking feature of the numerical study which clearly warrants further investigation is the fact that, when vaccine coverage is insufficient to prevent a major outbreak, the ordering of the performance of the household-based allocation strategies can be different depending on whether $R_v$ or $z$ is used as the measure of performance.

We also note in closing that there are some numerical challenges involved in implementing the methods we have presented. These are particularly relevant for the calculations relating to the forward process (i.e.\ calculation of $\pmaj$), but do also apply to the backward process (calculation of $z$). The main issue that arises is that of slowly converging infinite sums and the resulting possibilities for numerical overflow and underflow. Writing doubly or triply-infinite sums as one infinite and one or two finite sums (for example $\sum_{i=0}^\infty \sum_{j=0}^\infty a_{i,j}= \sum_{k=0}^\infty \sum_{l=0}^k a_{k-l,l}$) helps in some regards (e.g.\ faster computing since there is only one truncation to have to control the error of) but hinders in others (e.g.\ slower computing since methods to avoid underflow and overflow errors become more involved). Also, the case $p_Sp_N=0$ can present division by zero issues if not considered carefully.

% APPENDIX

\appendix

\section{APPENDIX}
\subsection{Notation}
\label{AppNotation}
The following notation is used throughout the appendix. The symbols $\mathbb{Z}_+$ and $\mathbb{R}_+$ denote the non-negative integers and reals, respectively. For vectors $\bs{x}=(x_1,x_2,\ldots,x_m)$ and $\bs{y}=(y_1,y_2,\ldots,y_m)$ in $\mathbb{R}^m$, we define $\bs{x}\bs{y}=\prod_{i=1}^m x_i y_i$ and $\bs{x}^{\bs{y}}=\prod_{i=1}^m x_i^{y_i}$.  We write $\bs{x}\le\bs{y}$ if $x_i \le y_i$ $(i=1,2,\ldots,m)$ and $\bs{x}<\bs{y}$ if, in addition, $x_i<y_i$ for at least one $i$.  For $n,k \in \mathbb{Z}_+$, the falling factorial $n!/(n-k)!$ is denoted by $n_{[k]}$.  For $\bs{n},\bs{k} \in \mathbb{Z}_+^m$, we define $\bs{n}_{[\bs{k}]}=\prod_{i=1}^m {n_i}_{[k_i]}$.  For $\bs{i},\bs{j} \in \mathbb{Z}_+^m$, we write $\sum_{\bs{k}=\bs{i}}^{\bs{j}}$ for $\sum_{k_1=i_1}^{j_1}\sum_{k_2=i_2}^{j_2}\ldots \sum_{k_m=i_m}^{j_m}$. Finally, we let $f^{(r)}$ denote the $r$-th order derivative of the function $f$.

\subsection{Multitype single-household epidemic and key result}
\label{AppKeyResult}
Consider a single-household SIR epidemic model with $m$ types of individuals, labelled $1,2,\ldots,m$.  Suppose that, for $i=1,2,\ldots,m$, there are initially $a_i$ infectives and $n_i$ susceptibles of type $i$, and let $\bs{a}=(a_1,a_2,\ldots,a_m)$ and $\bs{n}=(n_1,n_2,\ldots,n_m)$.  For $i=1,2,\ldots,m$, the infectious periods of type-$i$ infectives are each distributed according to a random variable $I^{(i)}$.  For $i,j=1,2,\ldots,m$, the individual-to-individual infection rate from a given type-$i$ infective to a given type-$j$ susceptible is $\lambda_{ij}$.  As in Section~\ref{subsecmod}, such infections are governed by Poisson processes, and all Poisson processes and infectious periods are mutually independent.

To each infective we attach a vector of $p$ non-negative integer-valued random attributes distributed according to a vector random variable
$\bs{A}^{(i)} = (A_1^{(i)}, A_2^{(i)}, \ldots , A_p^{(i)})$, where $i$ denotes the type of the infective.  In our applications $\bs{A}^{(i)}$ will describe the numbers of global infections of different types of individuals made by an infective in the single-household epidemic.  The realisations of the random variables $(I^{(i)}, \bs{A}^{(i)})$ are independent for distinct infectives and identically distributed for infectives of the same type.  Note that $I^{(i)}$ and $\bs{A}^{(i)}$ may be dependent.  For $i=1,2,\ldots,m$, let $T_i$ be the number of initial susceptibles of type $i$ that are ultimately infected by the epidemic and let $\bs{A}^{(i)} (T_i)$ be the sum of the attribute vectors over all $a_i + T_i$ infectives of type $i$.  Further, let
$\bs{T}=(T_1,T_2,\ldots,T_m)$ and let $\bs{A}(\bs{T})=\sum_{i=1}^m \bs{A}^{(i)} (T_i)$ be the sum of the attribute vectors over all
$\sum_{i=1}^m (a_i+T_i)$ infectives in the epidemic.  Thus, in applications, $\bs{A}(\bs{T})$ will give the offspring random variable for the forward branching process.  For $\bs{x}=(x_1,x_2,\ldots,x_m) \in \mathbb{R}^m$ and $\bs{s}=(s_1,s_2,\ldots,s_p) \in [0,1]^p$, let
\[
\phi (\bs{x},\bs{s}) = E[\bs{x}^{\bs{n}-\bs{T}} \bs{s}^{\bs{A}(\bs{T})}].
\]

We give below an expression for $\phi (\bs{x}, s)$ in terms of multivariate Gontcharoff polynomials, first studied by \citeA{LefPic1990}, which we now define.  Let $\bs{U}=(\bs{u}_{\bs{j}} \in \mathbb{R}^m \colon \bs{j} \in \mathbb{Z}_+^m$) be a collection of real numbers.  The Gontcharoff polynomials associated with $\bs{U}$, denoted $(G_{\bs{k}}(\bs{x} | \bs{U}) ,\, \bs{k}\in\mathbb{Z}_+^m , \bs{x}\in\mathbb{R}^m)$, are defined recursively by
\begin{equation}
\sum_{\bs{j}=\bs{0}}^{\bs{k}} \bs{k}_{[\bs{j}]} \bs{u}_{\bs{j}}^{\bs{k}-\bs{j}} G_{\bs{j}} (\bs{x} | \bs{U})=\bs{x}^{\bs{k}} \quad
(\bs{k} \geq \bs{0}).\label{GONT}
\end{equation}
Note that $G_{\bs{0}} (\bs{x}|\bs{U}) \equiv 1$ and that $G_{\bs{k}} (\bs{x}|\bs{U})$ is a polynomial of degree $k_1,k_2,\ldots,k_m$ in
$x_1,x_2,\ldots,x_m$, respectively, depending only on $(\bs{u}_{\bs{j}} \colon \bs{j} < \bs{k})$.  The following result is derived easily from \citeA[Theorem 5.1]{BalONe1999}, so its proof is omitted.

\begin{thm}
\label{thmBallONeill}
For $\bs{x} \in \mathbb{R}^m$ and $\bs{s} \in [0,1]^p$,
\begin{equation}
\phi (\bs{x}, \bs{s})=\sum_{\bs{j}=\bs{0}}^{\bs{n}} \bs{n}_{[\bs{j}]} (\bs{\psi}(\bs{s},\bs{j}))^{\bs{n}+\bs{a}-\bs{j}} G_{\bs{j}} (\bs{x}|\bs{U}),\label{PHIxs}
\end{equation}
where $\bs{\psi}(\bs{s},\bs{j})=(\psi_1 (\bs{s},\bs{j}), \psi_2 (\bs{s},\bs{j}), \ldots, \psi_m (\bs{s},\bs{j}))$, with
\begin{equation*}
\psi_i (\bs{s},\bs{j}) = E \left[ \exp \left( -I^{(i)} \sum_{k=1}^m \lambda_{ik} j_k \right) \bs{s}^{\bs{A}^{(i)}} \right] \quad (i=1,2,\ldots,m),
\end{equation*}
and $\bs{U} = (\bs{u}_{\bs{j}} \colon \bs{j} \in \mathbb{Z}_+^m)$ has components $\bs{u}_{\bs{j}} = \bs{\psi} (\bs{s},\bs{j})$.
\end{thm}

We now describe how Theorem~\ref{thmBallONeill} can be used to determine offspring PGFs for the forward branching processes in the main body of the paper.  Recall that primary and secondary infectives in a household typically have distinct global degree distributions.  Hence, in addition to being typed according to their vaccination status, individuals also need to be typed as primary or secondary.  Thus we may write $m=m_P+m_S$, where types $1,2,\ldots,m_P$ correspond to primary individuals and types $m_P +1, m_P +2, \ldots, m$ to secondary individuals.  Write $\bs{a}=(\bs{a}_P, \bs{a}_S)$,
$\bs{n}=(\bs{n}_P, \bs{n}_S)$, $\bs{x}=(\bs{x}_P, \bs{x}_S)$, $\bs{T}=(\bs{T}_P, \bs{T}_S)$ and $\bs{\psi}(\bs{s},\bs{j})=(\bs{\psi}_P (\bs{s},\bs{j}),
\bs{\psi}_S (\bs{s},\bs{j}))$ in the obvious fashion.  Note that all susceptibles are secondary individuals, so $\bs{n}_P=\bs{0}$, and all initial infectives are primary individuals, so $\bs{a}_S=\bs{0}$.  It follows that $\bs{T}_P=\bs{0}$ and the index $\bs{j}$ of the summation in
(\ref{PHIxs}) takes the form $(\bs{0},\bs{j}_S)$. Let
\[
\tilde{\phi} (\bs{x}_S, \bs{s}) = E[\bs{x}_S^{\bs{n}_S - \bs{T}_S} \bs{s}^{\bs{A}(\bs{T})} ] \quad (\bs{x}_S \in \mathbb{R}^{m_S}, \, \bs{s}\in [0,1]^p ).
\]
Then the following corollary follows easily from Theorem~\ref{thmBallONeill}.  For $i=1,2,\ldots,m$, let $\bs{\lambda}_S^{(i)}=(\lambda_{i,m_P+1},
\lambda_{i,m_P+2}, \ldots, \lambda_{i,m})^{\top}$, where $\top$ denotes transpose.

\begin{cor}
\label{corPrimSec}
For $\bs{x}_S \in \mathbb{R}^{m_S}$ and $\bs{s} \in [0,1]^p$,
\begin{equation}
\tilde{\phi} (\bs{x}_S, \bs{s}) = \sum_{\bs{j}_S=\bs{0}}^{\bs{n}_S} \bs{n}_{S_{[\bs{j}_S]}}
(\tilde{\bs{\psi}}_P (\bs{s},\bs{j}_S))^{\bs{a}_P} (\tilde{\bs{\psi}}_S (\bs{s},\bs{j}_S))^{\bs{n}_S-\bs{j}_S}
\,
G_{\bs{j}_S} (\bs{x}_S | \bs{U}_S),\label{PHITILDExs}
\end{equation}
where
$\tilde{\bs{\psi}}_P (\bs{s},\bs{j}_S)=\left(\tilde{\psi}_1 (\bs{s},\bs{j}_S), \tilde{\psi}_2 (\bs{s},\bs{j}_S), \ldots, \tilde{\psi}_{m_P} (\bs{s},\bs{j}_S)\right)$ \newline and
$\tilde{\bs{\psi}}_S (\bs{s}, \bs{j}_S) = \left(\tilde{\psi}_{m_P+1} (\bs{s},\bs{j}_S), \tilde{\psi}_{m_P+2} (\bs{s},\bs{j}_S), \ldots ,
\tilde{\psi}_m (\bs{s},\bs{j}_S)\right)$, with
\begin{equation}
\tilde{\psi}_i (\bs{s},\bs{j}_S)=E\left[\exp(-I^{(i)} \bs{j}_S \bs{\lambda}_S^{(i)}) \bs{s}^{\bs{A}^{(i)}}\right] \quad (i=1,2,\ldots,m)\label{PSITILDEsj}
\end{equation}
and $\bs{U}_S = ( \bs{u}_{\bs{j}_S}^S \colon \bs{j}_S \in \mathbb{Z}_+^{m_S})$ has components $\bs{u}_{\bs{j}_S}^S = \tilde{\bs{\psi}}_S (\bs{s},\bs{j}_S)$.
\end{cor}

We are primarily concerned with determining the PGF of $A(\bs{T})$, which of course is obtained by setting $\bs{x}_S=\bs{1}$ in (\ref{PHITILDExs}).
Thus, all that remains is to determine $\tilde{\psi}_i (\bs{s},\bs{j}_S)$, which is application dependent.  Note from (\ref{PSITILDEsj}) that it is sufficient to determine, for $\theta \in \mathbb{R}_+$ and $\bs{s} \in [0,1]^p$,
\[
\hat{\psi}_i (\theta,\bs{s}) = E\left[\exp (-\theta I^{(i)}) \bs{s} ^{\bs{A}^{(i)}}\right] \quad (i=1,2,\ldots,m),
\]
which we now do for the various models in the paper.

\subsection{No vaccination}
\label{AppNoVac}
This model is studied in \citeA{BalSirTra2010}, so we just state the result.  Note that there are two types of individual, primary ($P$) and secondary ($S$), with global forward degrees distributed according to $D_P \sim \tilde{D}-1$ and $D_S \sim D$, respectively.  Further $p=1$, so
$\bs{s}$ is a scalar, $s$ say.  For a degree distribution $D$ and real numbers $c_1,c_2,\theta$, define the function $F(D,c_1,c_2,\theta)$ by
\begin{equation}
F(D,c_1,c_2,\theta)=\sum_{r=0}^{\infty} \frac{c_1^r}{r!} \phi_I (\theta + r \lambda_G) f_D^{(r)} (c_2).\label{functionF}
\end{equation}
Then, for $A \in \{ P,S \}$,
\begin{equation*}
\hat{\psi}_A (\theta,s) = F(D_A, 1-s, s, \theta ),
\end{equation*}
cf.\ \citeA[Theorem 1]{BalSirTra2010}. Equation (\ref{functionF}), in conjunction with Corollary~\ref{corPrimSec}, enables the PGFs $f_{C^{(n)}}$ and $f_{\tilde{C}^{(n)}}$, defined in Section~\ref{subsecearlystages}, to be evaluated.

%Calculation of the mean within-household epidemic final size $\mu_n(\lambda_L)$ is addressed in Section~\ref{AppMeanLocalFS}.

\subsection{Households based vaccination}
\label{AppHouseVac}
Recall from Section~\ref{sechousevac} that with an all-or-nothing vaccine action all required PGFs and means can be expressed in terms of $f_{C^{(n)}}$, $f_{\tilde{C}^{(n)}}$ and $\mu_n (\lambda_L)$, so here we need consider only the non-random vaccine action model.  This model has four types of individuals: primary-unvaccinated ($PU$), primary-vaccinated ($PV$), secondary-unvaccinated ($SU$) and secondary-vaccinated ($SV$).  Denote their respective forward global degree distributions by $D_{PU}$, $D_{PV}$, $D_{SU}$ and $D_{SV}$, respectively, and recall from Section~\ref{subsechousevacnr} that $D_{PU} \stackrel{D}{=} D_{PV} \sim \tilde{D}-1$ and $D_{SU} \stackrel{D}{=} D_{SV} \sim D$.  For $F \in \{PU,PV,SU,SV\}$, the random attribute of interest is
$\bs{A}^{(F)}=(A_U^{(F)},A_V^{(F)})$, where $A_U^{(F)}$ and $A_V^{(F)}$ are the number of unvaccinated and vaccinated global infections made by a type-$F$ infective in the single household epidemic.  Thus $m_S=2$ and we determine $\hat{\psi}_F (\theta,\bs{s})$, where $\bs{s}=(s_U,s_V)$. 

Consider first $\hat{\psi}_{SU}(\theta,\bs{s})$.  Conditioning on the degree $D_{SU}$ and infectious period $I$ of a typical type-$SU$ infective, $i^{\ast}$ say, yields
\begin{equation}
\hat{\psi}_{SU} (\theta,\bs{s}) = E_{D_{SU},I} \left[ \re^{-\theta I} E\left[\bs{s}^{\bs{A}^{(SU)}} \mid D_{SU}, I\right]\right].\label{PSIHATSU}
\end{equation}
Let $N_V$ denote the number of $i^{\ast}$'s global neighbours that are vaccinated.  Then \newline
$\left(A_U^{(SU)} \mid N_V, D_{SU}, I\right) \sim \textrm{Bin} \left(D_{SU} - N_V, 1-\re^{-\lambda_G I}\right)$ and \newline
$\left(A_V^{(SU)} \mid N_V, D_{SU}, I\right) \sim \textrm{Bin} \left(N_V, 1-\re^{-\lambda_G aI}\right)$ independently, so
\begin{equation}
E\left[\bs{s}^{\bs{A}^{(SU)}} \mid N_V, D_{SU}, I\right]=\left(\re^{-\lambda_G I} + (1-\re^{-\lambda_G I})s_U \right)^{D_{SU}-N_V}
\left(\re^{-\lambda_G aI} + (1-\re^{-\lambda_G aI})s_V\right)^{N_V},\label{EsANDI}
\end{equation}
whence, since $(N_V \mid D_{SU}) \sim \textrm{Bin} (D_{SU},p_V)$,
\begin{equation}
E\left[\bs{s}^{\bs{A}^{(SU)}} \mid D_{SU}, I\right] =
\left((1-p_V) \left[ \re^{-\lambda_G I} + (1- \re^{-\lambda_G I}) s_U \right] + p_V \left[ \re^{-\lambda_G aI} + (1-\re^{-\lambda_G aI})s_V \right]\right)^{D_{SU}}.
\label{EsADI}
\end{equation}
(As usual, $\textrm{Bin}(n,p)$ denotes a binomial distribution with parameters $n\in\mathbb{Z}_+$ and $p\in[0,1]$.)
Let $\hat{a} (\bs{s}) = p_V s_V + (1-p_V) s_U$,
$\hat{b} (\bs{s}) = p_V (1-s_V)$ and
$\hat{c} (\bs{s}) = (1-p_V)(1-s_U)$.  Then, using (\ref{PSIHATSU}) and (\ref{EsADI}),
\begin{align}
\hat{\psi}_{SU} (\theta,\bs{s})
&=E_{D_{SU},I} \left[ \re^{-\theta I} \left(\hat{a} (\bs{s}) + \hat{b} (\bs{s}) \re^{-\lambda_G aI} + \hat{c} (\bs{s}) \re^{-\lambda_G I}\right)^{D_{SU}}\right]\nonumber\\
&=\sum_{k=0}^{\infty} P(D_{SU}=k) E_I \Bigg[ \re^{-\theta I} \sum_{r=0}^k \sum_{l=0}^{k-r} \frac{k!}{r! l! (k-r-l)!} (\hat{a} (\bs{s}))^{k-l-r}
 \nonumber\\
& \hspace{6.5cm} \times  (\hat{b} (\bs{s}) \re^{-\lambda_G aI})^r (\hat{c} (\bs{s}) \re^{-\lambda_G I})^l \Bigg] \nonumber\\
&=\sum_{r=0}^\infty \sum_{l=0}^\infty \sum_{k=r+l}^\infty P(D_{SU} = k) \frac{k!}{r! l! (k-r-l)!} (\hat{a} (\bs{s}))^{k-l-r} \nonumber\\
& \hspace{5cm} \times (\hat{b} (\bs{s}))^r (\hat{c} (\bs{s}))^l E\left[\re^{-(\theta + (ar+l) \lambda_G)I}\right].
\label{PSIHATSUSuM}
\end{align}

For a degree distribution $D$ and real numbers $c_1,c_2,c_3,\theta,a,b$, define the function \newline $G(D,c_1,c_2,c_3,\theta,a,b)$ by
\begin{equation}
G(D,c_1,c_2,c_3,\theta,a,b)=\sum_{r=0}^\infty \sum_{l=0}^\infty \frac{c_2^r}{r!} \frac{c_3^l}{l!} \phi_I (\theta + \lambda_G b(ar+l)) f_D^{(r+l)} (c_1).\label{functionG}
\end{equation}
Then (\ref{PSIHATSUSuM}) implies that
\begin{equation}
\hat{\psi}_{SU} (\theta,\bs{s})=G(D_{SU}, \hat{a}(\bs{s}),\hat{b}(\bs{s}),\hat{c}(\bs{s}),\theta,a,1).
\label{PSIHATSUG}
\end{equation}
A similar argument shows that
\begin{equation}
\hat{\psi}_{SV} (\theta,\bs{s}) = G(D_{SV}, \hat{a}(\bs{s}), \hat{b}(\bs{s}), \hat{c}(\bs{s}), \theta, a,b),
\label{PSIHATSVG}
\end{equation}
and that $\hat{\psi}_{PU} (\theta,\bs{s})$ is given by (\ref{PSIHATSUG}), with $D_{SU}$ replaced by $D_{PU}$, and $\hat{\psi}_{PV} (\theta,\bs{s})$ is given by (\ref{PSIHATSVG}), with $D_{SV}$ replaced by $D_{PV}$.

If a closed-form expression for $f_D$ is available then $G(D,c_1,c_2,c_3,\theta,a,b)$ may be computed by using a finite truncation of
(\ref{functionG}).  If a closed-form expression for $f_D$ is unavailable then $G(D,c_1,c_2,c_3,\theta,a,b)$ may be computed by using a finite truncation of a corresponding triple sum (cf.\ (\ref{PSIHATSUSuM})).

Note that if the infectious period is constant, say $I \equiv \iota$, then (\ref{PSIHATSU}) to (\ref{EsADI}) imply that
\begin{equation*}
\hat{\psi}_{SU} (\theta,\bs{s}) = \re^{-\theta \iota} f_{D_{SU}} \left(\hat{a}(\bs{s}) + \re^{-\lambda_G a \iota} \hat{b}(\bs{s}) + \re^{-\lambda_G \iota} \hat{c}(\bs{s})\right)
\end{equation*}
and corresponding expressions for $\hat{\psi}_{SV} (\theta,\bs{s})$, $\hat{\psi}_{PU} (\theta,\bs{s})$ and
$\hat{\psi}_{PV} (\theta,\bs{s})$ are easily derived.

%Formulae for the means $\mu^{(n,v)} (A,A')$ ($A,A' \in \{U,V\}$) (see Section~\ref{subsechousevacnr}) are given in Section~\ref{AppMeanLocalFS}.

\subsection{Acquaintance vaccination}
\label{AppAcqVac}
This model has eight types of individuals, six primary and two secondary, which, as in Section~\ref{subsecacqmod}, we label $1,2,\ldots,6$ and $U,V$, respectively.  For each type $i$, the random attribute of interest $\bs{A}^{(i)}=(A^{(i)}_1,A^{(i)}_2,\ldots,A^{(i)}_6)$, where $A^{(i)}_j$ is the number of global infections of a type $j$ individual made by an infective type $i$ individual.  Thus we determine, in an obvious notation, $\hat{\psi}_i(\theta,\bs{s})$ ($i=1,2,\ldots,6$), $\hat{\psi}_U (\theta,\bs{s})$ and $\hat{\psi}_V (\theta,\bs{s})$; for $\theta\in\mathbb{R}_+$ and $\bs{s}=(s_1,s_2,\ldots,s_6) \in [0,1]^6$.  It is convenient to treat the all-or-nothing and non-random vaccine action models separately.

\subsubsection{All-or-nothing vaccine action}
\label{AppAcqAoN}
For $i=1,2,\ldots,6$, recall that $\tilde{D}(i)$ is the global degree of a typical type-$i$ primary individual, $i^{\ast}$ say, and let
$\tilde{\bs{X}}_i = (\tilde{X}_{i1}, \tilde{X}_{i2}, \ldots, \tilde{X}_{i6})$, where $\tilde{X}_{ij}$ is the number of $\bs{i}^{\ast}$'s forward global neighbours that have type $i$.  (We use the same notation as for the backward process in Section~\ref{subsecacqfinal}, since the distribution of
$(\tilde{D}(i),\tilde{\bs{X}}_i)$ is the same for the forward and backward processes.)  Recall that types 3 and 6 have not been vaccinated and types 1, 2, 4 and 5 have been vaccinated.  Thus
\begin{equation*}
\left(A_j^{(i)}| \tilde{X}_{ij},I\right) \sim \left\{
\begin{array}{ll}\textrm{Bin}\left(\tilde{X}_{ij}, 1-\re^{-\lambda_GI}\right)&\text{if }j=3,6,\\
\textrm{Bin}\left(\tilde{X}_{ij}, (1-\epsilon)(1-\re^{-\lambda_GI})\right)&\text{if } j=1,2,4,5,
\end{array} \right.
\end{equation*}
and
\begin{align}
E\left[\bs{s}^{\bs{A}^{(i)}} | \tilde{\bs{X}}_i,I\right]=\prod_{j=3,6} &\left(\re^{-\lambda_GI} +  \left. (1-\re^{-\lambda_GI})s_j\right)^{\tilde{X}_{ij}}\right.\nonumber\\
& \times \prod_{j=1,2,4,5} \left( \epsilon + (1- \epsilon ) \re^{-\lambda_GI} + (1-\epsilon)(1-\re^{-\lambda_G I} ) s_j\right)^{\tilde{X}_{ij}}.
\label{PGFAXI}
\end{align}
Further, for $i=1,3,4,6$, the types of the $\tilde{D}(i)-1$ of $i^{\ast}$'s forward global neighbours are chosen independently according to
$\hat{p}_{ij}$, defined in Section~\ref{subsecacqthreshold}, so
\begin{equation}
E\left[\bs{s}^{\bs{A}^{(i)}} | \tilde{D} (i),I\right] = \left(\hat{p}_i (\bs{s},\epsilon) + (1-\hat{p}_i (\bs{s},\epsilon)) \re^{-\lambda_GI}\right)^{\tilde{D}(i)-1},
\label{PGFADI}
\end{equation}
where
\begin{equation*}
\hat{p}_i (\bs{s},\epsilon) = \sum_{j=3,6} \hat{p}_{ij} s_j + \sum_{j=1,2,4,5} \hat{p}_{ij} (\epsilon + (1-\epsilon) s_j).
\end{equation*}
Hence,
\begin{align}
\hat{\psi}_i (\theta,\bs{s})&=E\left[\re^{-\theta I} \bs{s}^{\bs{A}^{(i)}}\right]\nonumber\\
&=E_I \left[\re^{-\theta I} E_{\tilde{D}(i)} \left[ E\left[ \bs{s}^{\bs{A}^{(i)}} | \tilde{D} (i),I \right] \right] \right] \nonumber\\
&=E_I \left[ \re^{-\theta I} E_{\tilde{D}(i)} \left[ ( \hat{p}_i (\bs{s},\epsilon) + (1-\hat{p}_i (\bs{s},\epsilon)) \re^{-\lambda_GI})^{\tilde{D}(i)-1} \right] \right].
\label{PSIHATi}
\end{align}
A similar argument to the derivation of (\ref{PSIHATSUSuM}), using the binomial theorem rather than the multinomial theorem, then yields
\begin{equation*}
\hat{\psi}_i (\theta,\bs{s}) = F(\tilde{D}(i)-1, 1-\hat{p}_i (\bs{s},\epsilon), \hat{p}_i (\bs{s},\epsilon), \theta ).
\end{equation*}

For $i=2$ and $i=5$, (\ref{PGFAXI}) still holds but $f_{\tilde{\bs{X}}_i} (\bs{s})$ is given by (\ref{fX2tilde}) and (\ref{fX5tilde}), respectively.
Omitting the details, similar arguments to the above show that, for $i=2,5$,
\begin{equation}
\hat{\psi}_i (\theta,\bs{s}) = \tilde{p}_V^{-1} \left[ F\left(\tilde{D}-1,\tilde{p}_i (\bs{s},0), \tilde{q}_i (\bs{s},0),\theta\right)
- F\left(\tilde{D}-1, \tilde{p}_i (\bs{s},p_N), \tilde{q}_i (\bs{s}, p_N ), \theta \right)\right],
\label{PSIHAT25}
\end{equation}
where $\tilde{p}_i(\bs{s},x) = p_S(1-x)\tilde{p}_{i1}(\bs{s}) + (1-p_S)\tilde{p}_{i2}(\bs{s})$ and $\tilde{q}_i(\bs{s},x) = 1-p_Sx-\tilde{p}_i(\bs{s},x)$, with
\begin{align*}
\tilde{p}_{21}(\bs{s}) & = p_N(1-\epsilon)(1-s_1) + (1-p_N)[\tilde{p}_V(1-\epsilon)(1-s_2) + (1-\tilde{p}_V)(1-s_3)], \\
\tilde{p}_{22}(\bs{s}) & = p_N(1-\epsilon)(1-s_4) + (1-p_N)[\tilde{p}_V(1-\epsilon)(1-s_5) + (1-\tilde{p}_V)(1-s_6)], \\
\tilde{p}_{51}(\bs{s}) & = \tilde{p}_V(1-\epsilon)(1-s_2) + (1-\tilde{p}_V)(1-s_3) \quad \mbox{and} \\
\tilde{p}_{52}(\bs{s}) & = \tilde{p}_V(1-\epsilon)(1-s_5) + (1-\tilde{p}_V)(1-s_6).
\end{align*}

To simplify the exposition we write (\ref{PSIHAT25}) as
\begin{equation*}
\hat{\psi}_i (\theta,\bs{s}) = \tilde{p}_V^{-1} \Delta_{0,p_N} F\left(\tilde{D}-1,\tilde{p}_i (\bs{s},x), \tilde{q}_i (\bs{s},x),\theta\right).
\end{equation*}
In the sequel we use, without comment, a similar notation for differences of other functions evaluated at $0$ and $p_N$.

Turning now to the secondary individuals, note that a type-$U$ individual is sampled with probability $p_S$, in which case it behaves like a type-3 primary individual but with forward global degree distributed according to $D_U$, otherwise it is unsampled and behaves like a type-6 primary individual, again with forward global degree distributed according to $D_U$.  Thus
\begin{equation}
\hat{\psi}_U (\theta,\bs{s}) = p_S F\left(D_U,1-\hat{p}_3 (\bs{s},\epsilon), \hat{p}_3 (\bs{s},\epsilon),\theta\right)+(1-p_S)F\left(D_U,1-\hat{p}_6 (\bs{s},\epsilon\right),
\hat{p}_6 (\bs{s},\epsilon),\theta).
\label{PSIUHAT}
\end{equation}
Similarly,
\begin{align}
\hat{\psi}_V (\theta,\bs{s})&=
p_V^{-1} [ p_S \Delta_{0,p_N} F\left(D_V, \tilde{p}_2 (\bs{s},x), \tilde{q}_2 (\bs{s},x),\theta\right) \nonumber \\
& \qquad + (1-p_S) \Delta_{0,p_N} F\left(D_V,\tilde{p}_5 (\bs{s},x), \tilde{q}_5 (\bs{s},x),\theta\right)].
\label{PSIVHAT}
\end{align}
To simplify the exposition we write (\ref{PSIUHAT}) and (\ref{PSIVHAT}) as
\begin{equation*}
\hat{\psi}_U (\theta,\bs{s})=p_S^{3,6} F(D_U,1-\hat{p}(\bs{s},\epsilon),\hat{p}(\bs{s},\epsilon),\theta)
\end{equation*}
and
\begin{equation*}
\hat{\psi}_V (\theta,\bs{s}) = p_V^{-1} p_S^{2,5} \Delta_{0,p_N} F(D_V,\tilde{p}(\bs{s},x), \tilde{q}(\bs{s},x),\theta).
\end{equation*}
In the sequel, we use a similar notation without comment.

The above expressions simplify appreciably if the infectious period is constant, say $I \equiv \iota$.  In that case, equation (\ref{PSIHATi}) implies that
\begin{equation*}
\hat{\psi}_i (\theta,\bs{s}) = \re^{-\theta \iota} f_{\tilde{D}(i)-1} \left(\hat{p}_i (\bs{s},\epsilon) + (1-\hat{p}_i (\bs{s},\epsilon)) \re^{-\lambda_G \iota}\right) \quad (i=1,3,4,6)
\end{equation*}
and exploiting the PGFs for $\tilde{\bs{X}}_2$ and $\tilde{\bs{X}}_5$ (see (\ref{fX2tilde}) and (\ref{fX5tilde})) yields
\begin{equation*}
\hat{\psi}_i (\theta,\bs{s}) = \tilde{p}_V^{-1} \re^{-\theta \iota} \Delta_{0,p_N} f_{\tilde{D}-1} \left(\tilde{q}_i (\bs{s},x) + \tilde{p}_i (\bs{s},x) \re^{-\lambda_G \iota}\right) \quad (i=2,5).
\end{equation*}
Similar arguments show that
\begin{equation*}
\hat{\psi}_U (\theta,\bs{s})= \re^{-\theta \iota} p_S^{3,6} f_{D_U} \left(\hat{p} (\bs{s},\epsilon)   + (1-\hat{p} (\bs{s},\epsilon)) \re^{-\lambda _G \iota}\right)
\end{equation*}
and
\begin{equation*}
\hat{\psi}_V (\theta,\bs{s}) = p_V^{-1} \re^{-\theta \iota} p_S^{2,5} \Delta_{0,p_N} f_{D_V} \left(\tilde{q} (\bs{s},x) + \tilde{p} (\bs{s},x)
\re^{-\lambda_G \iota} \right).
\end{equation*}

\subsubsection{Non-random vaccine}
\label{AppAcqNR}
Note that now
\begin{equation}
(A_j^{(i)} | \tilde{X}_{ij},I) \sim \left\{
\begin{array}{ll}
\textrm{Bin} ( \tilde{X}_{ij}, 1-\re^{-\lambda _G I})&\text{if } i,j \in \{3,6\},\\
\textrm{Bin} (\tilde{X}_{ij}, 1-\re^{-a \lambda _G I} )&\text{if } i \in \{3,6\}, ~ j \in \{ 1,2,4,5),\\
\textrm{Bin} ( \tilde{X}_{ij}, 1-\re^{-b \lambda_G I})&\text{if } i \in \{ 1,2,4;5\}, ~ j \in \{ 3,6 \},\\
\textrm{Bin} ( \tilde{X}_{ij}, 1-\re^{-ab \lambda_G I})&\text{if } i,j \in \{ 1,2,4,5 \}.
\end{array} \right.
\label{AXINR}
\end{equation}
For $i \in \{1, 3,4,6 \}$, let
\begin{equation*}
\hat{a}_i (\bs{s}) = \sum_{j=1}^6 \hat{p}_{ij} s_j, \quad
\hat{b}_i (\bs{s}) = \sum_{j=1,2,4,5} \hat{p}_{ij} (1-s_j) \quad \text{and} \quad
\hat{c}_i (\bs{s}) = \sum_{j=3,6} \hat{p}_{ij} (1-s_j).
\end{equation*}
Suppose that $i \in \{ 3,6 \}$.  Then,  arguing as in the derivation of (\ref{PGFADI}),
\begin{equation*}
E\left[\bs{s}^{\bs{A}^{(i)}} | \tilde{D}(i),I\right] = \left(\hat{a}_i (\bs{s}) + \hat{b}_i (\bs{s}) \re^{-a \lambda_G I} + \hat{c}_i (\bs{s}) \re^{-\lambda_G I}\right)^{\tilde{D} (i)-1} ,
\end{equation*}
whence
\begin{equation}
\hat{\psi}_i (\theta,\bs{s}) = E_I \left[ \re^{- \theta I} E_{\tilde{D}(i)} \left[\left(\hat{a}_i (\bs{s}) + \hat{b}_i (\bs{s}) \re^{-a \lambda_G I} +
\hat{c}_i (\bs{s}) \re^{-\lambda_G I} \right) ^{\tilde{D}(i)-1} \right]\right].
\label{PSIHATiNR}
\end{equation}
Arguing as in the derivation of (\ref{PSIHATSUSuM}) then yields that
\begin{equation*}
\hat{\psi}_i (\theta,\bs{s}) = G\left(\tilde{D}(i)-1, \hat{a}_i (\bs{s}), \hat{b}_i (\bs{s}), \hat{c}_i (\bs{s}), \theta , a, 1\right).
\end{equation*}
Similarly, for $i \in \{ 1,4 \}$,
\begin{equation*}
\hat{\psi}_i (\theta,\bs{s}) = G\left(\tilde{D}(i) -1, \hat{a}_i (\bs{s}), \hat{b}_i (\bs{s}), \hat{c}_i (\bs{s}), \theta, a,b\right).
\end{equation*}

Suppose that $i=5$.  Then, using (\ref{AXINR}),
\begin{equation*}
E\left[\bs{s}^{\bs{A}^{(5)}} | \tilde{\bs{X}}_5, I\right] = \prod_{j=3,6} \left( \re^{-b \lambda_G I} + (1-\re^{-b \lambda_GI})s_j\right)^{\tilde{X}_{5j}}
\prod_{j=1,2,4,5} \left(\re^{-ab \lambda_G I} + (1-\re^{-ab \lambda_G I} ) s_j \right)^{\tilde{X}_{5j}}.
\end{equation*}
Invoking (\ref{fX5tilde}) now gives
\begin{equation}
\hat{\psi}_5 (\theta,\bs{s}) = \tilde{p}_V^{-1} E\left[\re^{-\theta I} \Delta_{0,p_N} f_{\tilde{D}-1} \left(\hat{g}_5 (\bs{h}_5^F (\bs{s},I),
x)\right)\right],
\label{PSI5HAT}
\end{equation}
where $\bs{h}_5^F (\bs{s},I) = (h_{51}^F (s_1,I), h_{52}^F (s_2,I), \ldots, h_{56}^F (s_6,I))$, with $h_{5j}^F(s,I)=\re^{-b \lambda_G I} +
(1-\re^{-b \lambda_G I})s$ if $j=3,6$ and $h_{5j}^F (s,I) = \re^{-ab \lambda_G I} + (1-\re^{-ab \lambda_G I})s$ if $j=1,2,4,5$; and $\hat{g}_5$ is as defined in Section~\ref{subsecacqfinal}.  A simple calculation now shows that
\begin{equation}
\hat{g}_5 \left(\bs{h}_5^F (\bs{s},I), x \right) = \hat{a}_5 (\bs{s},x) + \hat{b}_5 (\bs{s},x) \re^{-ab \lambda_G I} +
\hat{c}_5 (\bs{s},x) \re^{-b \lambda_G I},
\label{g5hat}
\end{equation}
where $\hat{a}_5 (\bs{s},x)=p_S(1-x)[\tilde{p}_V s_2 + (1-\tilde{p}_V)s_3] + (1-p_S)[\tilde{p}_V s_5 + (1-\tilde{p}_V)s_6]$, 
$\hat{b}_5 (\bs{s},x)=p_S(1-x) \tilde{p}_V (1-s_2) + (1-p_S) \tilde{p}_V (1-s_5)$ and
$\hat{c}_5 (\bs{s},x)=p_S(1-x)(1-\tilde{p}_V)(1-s_3) + (1-p_S)(1-\tilde{p}_V)(1-s_6)$.  Arguing as in the derivation of (\ref{PSIHATSUSuM}) now gives, for $i=5$,
\begin{equation}
\hat{\psi}_i (\theta,\bs{s}) = \tilde{p}_V^{-1} \Delta_{0,p_N} G\left(\tilde{D}-1, \hat{a}_i (\bs{s},x),\hat{b}_i (\bs{s},x), \hat{c}_i (\bs{s},x), \theta, a,b\right).
\label{PSI2HAT}
\end{equation}
A similar argument shows that (\ref{PSI2HAT}) holds also for $i=2$, with
$\hat{a}_2 (\bs{s},x)=p_S(1-x) \{ p_N s_1 + (1-p_N) [ \tilde{p}_V s_2 + (1-\tilde{p}_V) s_3]\} + 
(1-p_S) \{ p_N s_4 + (1-p_N) [ \tilde{p}_V s_5 + (1-\tilde{p}_V) s_6 ] \}$,
$\hat{b}_2 (\bs{s},x) = p_S (1-x) [p_N (1-s_1) + (1-p_N) \tilde{p}_V (1-s_2)] + (1-p_S) [p_N (1-s_4) + (1-p_N) \tilde{p}_V (1-s_5)]$ and
$\hat{c}_2 (\bs{s},x) = (1-p_N)(1-\tilde{p}_V) [p_S (1-x)(1-s_3) + (1-p_S)(1-s_6)]$.

Expressions for $\hat{\psi}_U (\theta,\bs{s})$ and $\hat{\psi}_V (\theta,\bs{s})$ are derived in exactly the same way as with an all-or-nothing vaccine, yielding
\[
\hat{\psi}_U (\theta,\bs{s})=p_S^{3,6} G\left(D_U , \hat{a} (\bs{s}), \hat{b} (\bs{s}), \hat{c} (\bs{s}), \theta , a, 1\right)
\]
and
\[
\hat{\psi}_V (\theta,\bs{s}) = p_V^{-1} p_S^{2,5} \Delta_{0,p_N} G\left(D_V, \hat{a}(\bs{s},x), \hat{b} (\bs{s},x), \hat{c}(\bs{s},x),\theta,a,b\right).
\]

As usual, the above expressions simplify if $I \equiv \iota$.  Equation (\ref{PSIHATiNR}) then implies that, for $i=3,6$,
\[
\hat{\psi}_i (\theta,\bs{s}) = \re^{-\theta \iota} f_{\tilde{D}(i)} \left(\hat{a}_i (\bs{s}) + \hat{b}_i (\bs{s}) \re^{-a \lambda_G \iota} + \hat{c}_i (\bs{s})
\re^{-\lambda_G \iota}\right),
\]
which also holds for $i=1,4$, provided $\lambda_G$ is replaced by $b \lambda_G$.  Further, exploiting (\ref{PSI5HAT}), (\ref{g5hat}) and similar equations for $\hat{\psi}_2 (\theta,\bs{s})$, gives, for $i=2,5$, 
\[
\hat{\psi}_i (\theta,\bs{s}) = \tilde{p}_V^{-1} \re^{-\theta \iota} \Delta_{0,p_N} f_{\tilde{D}-1} \left(\hat{a}_i (\bs{s},x) + \hat{b}_i (\bs{s},x) \re^{-ab \lambda_G \iota} + \hat{c}_i (\bs{s},x) \re^{-b \lambda_G \iota}\right).
\]
Further,
\[
\hat{\psi}_U (\theta,\bs{s}) = \re^{-\theta \iota} p_S^{3,6} f_{D_U} \left(\hat{a}(\bs{s}) + \hat{b}(\bs{s}) \re^{-a \lambda_G \iota} + \hat{c}(\bs{s}) \re^{-\lambda_G \iota}\right)
\]
and
\[
\hat{\psi}_V (\theta,s) = p_V^{-1} \re^{-\theta \iota} p_S^{2,5} \Delta_{0,p_N} f_{D_V} \left(\hat{a}(\bs{s},x) + \hat{b}(\bs{s},x) \re^{-ab \lambda_G \iota} + \hat{c} (\bs{s},x) \re^{-b \lambda_G \iota}\right).
\]

\subsection{Local susceptibility set size}
\label{AppSusset}
In this section we give formulae for the probability mass functions of local susceptibility sets.  Consider first the model without vaccination and recall the definition of $M^{(n)}$ in Section~\ref{subsecfinaloutcome}.  Then it follows directly from \citeA[Lemma~3.1]{Ball2000ResRep} (see also \citeA[Lemma~3.1]{BalNea2002}, which gives the same result but not in terms of Gontcharoff polynomials) that 
\begin{equation}
P(M^{(n)} = k) = (n-1)_{[k]} q_{k+1}^{n-1-k} G_{k+1} (1|V) \quad (k=0,1,\ldots,n-1),\label{susmass1}
\end{equation}
where $q_i = \phi_I (\lambda_L i)$ and $V=(q_{i+1} \colon i=0,1,\ldots )$.

Consider next the model with a non-random vaccine action and recall the definition of $\bs{M}_A^{(n,v)} = (M_{AU}^{(n,v)}, M_{AV}^{(n,v)})$ ($A \in \{ U,V \}$) in Section~\ref{subsechousevacnr}.  It is convenient to now use a different notation.  For $\bs{n}=(n_U,n_V) \in \mathbb{Z}_+^2$, let $\hat{\bs{M}}_U^{(\bs{n})}=(\hat{M}_{UU}^{(\bs{n})}, \hat{M}_{UV}^{(\bs{n})})$, where $\hat{M}_{UU}^{(\bs{n})}$ and $\hat{M}_{UV}^{(\bs{n})}$ are the numbers of unvaccinated and vaccinated individuals, respectively, in the local susceptibility set of an unvaccinated individual, $i^{\ast}$ say, who resides in a household containing $n_U$ other unvaccinated individuals and $n_V$ vaccinated individuals, so the household has size $n_U + n_V + 1$.  Note that $\hat{M}_{UU}^{(\bs{n})}$ does \emph{not} include the individual $i^{\ast}$.  Define $\hat{\bs{M}}_V^{(\bs{n})} = (\hat{M}_{VU}^{(\bs{n})}, \hat{M}_{VV}^{(\bs{n})})$ similarly, but for a vaccinated individual. Hence, to connect with the notation in Section~\ref{subsechousevacnr}, $\bs{M}_U^{(n,v)}=\hat{\bs{M}}_U^{(n-1-v,v)}$ and $\bs{M}_V^{(n,v)}=\hat{\bs{M}}_V^{(n-v,v-1)}$.  For $\bs{n} \geq \bs{0}$ and $A \in \{ U,V \}$, the probability mass function of $\hat{\bs{M}}_A^{(\bs{n})}$ can be derived using the same argument as in the proof of Lemma~3.1 in \citeA{Ball2000ResRep}.  For brevity, we omit the details and just state the result.

For $\bs{j}=(j_U,j_V) \geq \bs{0}$, let $\bs{q}_{\bs{j}} = (q_{\bs{j}}^U, q_{\bs{j}}^V)$, where $q_{\bs{j}}^U = \phi_I (\lambda_L (j_U+bj_V))$ and $q_{\bs{j}}^V = \phi_I (a\lambda_L (j_U+bj_V))$.  Let $\bs{1}_U = (1,0)$ and $\bs{1}_V=(0,1)$.  Then, for $\bs{n} \geq \bs{0}$ and $A \in \{ U,V \}$, 
\begin{equation}
P(\hat{\bs{M}}_A^{(\bs{n})} = \bs{k}) = \bs{n}_{[\bs{k}]} \bs{q}_{\bs{k}+\bs{1}_A}^{\bs{n}-\bs{k}} G_{\bs{k}} (\bs{1} | \bs{V}^A) \quad (\bs{0} \leq \bs{k} \leq \bs{n}-\bs{1}_A),\label{susmassNR}
\end{equation}
where $\bs{V}^A = (\bs{v}_{\bs{j}}^A \colon \bs{j} \geq \bs{0})$ has components $\bs{v}_{\bs{j}}^A = \bs{q}_{\bs{j}+\bs{1}_A}^A$.

In the model with an all-or-nothing vaccine action we can calculate the mass function of $\bs{M}_A^{(n,v)}$ by conditioning on the number, $v_S$ say, of the $v$ vaccinations that are successful. The mass function of $\bs{M}_A^{(n-v_S,v-v_S)}$ can then be calculated as above, but with $q_{\bs{j}}^U = q_{\bs{j}}^V = \phi_I (\lambda_L (j_U+j_V))$.

\subsection{Mean local epidemic size}
\label{AppMeanLocalFS}

The mean single-household epidemic final size $\mu_n (\lambda_L)$, defined in Section~\ref{subsecearlystages}, is given in terms of Gontcharoff polynomials by
\begin{equation}
\mu_n (\lambda_L) = n-1- \sum_{i=1}^{n-1} (n-1)_{[i]} q_i^{n-i} G_{i-1} (1|V),\label{munlambdaL}
\end{equation}
where $q_i= \phi_I (i \lambda_L )$ and $V=(q_{i+1} \colon i=0,1,\ldots )$; cf.\ \citeA[Corollary~3.3]{LefPic1990}.

Finally, we consider the means $\mu^{(n,v)} (A,A')$ ($A,A' \in \{ U,V \}$) defined in Section~\ref{subsechousevacnr}. Again it is convenient to use a different notation.  For $\bs{n}=(n_U,n_V) \geq \bs{0}$, let $\hat{\mu}^{(\bs{n})} (U,U)$ and $\hat{\mu}^{(\bs{n})} (U,V)$ denote respectively the mean number of unvaccinated and vaccinated susceptibles that are ultimately infected in a single-household epidemic with 1 initial infective, who is unvaccinated, $n_U$ unvaccinated susceptibles and $n_V$ vaccinated susceptibles.  Define $\hat{\mu}^{(\bs{n})} (V,U)$ and $\hat{\mu}^{(\bs{n})} (V,V)$ similarly, for when the initial infective is vaccinated.  Thus, to connect with the notation in Section~\ref{subsechousevacnr}, for $A \in \{ U,V \}$, $\mu^{(n,v)} (U,A)=\hat{\mu}^{(n-1-v,v)} (U,A)$ and $\mu^{(n,v)} (V,A) = \hat{\mu}^{(n-v,v-1)} (V,A)$.  Suitable differentiation applied to Theorem~3.5 of \citeA{Ball1986}, or Corollary~4.4 of \citeA{PicLef1990}, yields that, for $\bs{n} \geq 0$ and $A,A' \in \{ U,V \}$, 
\begin{equation}
\hat{\mu}^{(\bs{n})} (A,A') = n_{A'} - \sum_{\bs{k}=\bs{1}_{A'}}^{\bs{n}} \bs{n}_{[\bs{k}]} 
\bs{q}_{\bs{k}}^{\bs{n}+\bs{1}_A - \bs{k}} G_{\bs{k}-\bs{1}_{A'}} (\bs{1} | \bs{V}^{A'}),\label{meanNR}
\end{equation}
where $\bs{q}_{\bs{k}}$ ($\bs{k} \geq \bs{0}$) and $\bs{V}^A$ ($A \in \{ U,V \}$) are as in (\ref{susmassNR}).

\section*{Acknowledgements}

This work was partially supported by a grant from the Simons Foundation. It was also supported by the UK Engineering and Physical Sciences Research Council (EPSRC), through research grant EP/E038670/1. The authors thank the Isaac Newton Institute for Mathematical Sciences, Cambridge, for support and hospitality during the programme Theoretical Foundations for Statistical Network Analysis, where work on this paper was undertaken (EPSRC grant EP/K032208/1).

\bibliography{bibfile}
\bibliographystyle{apacite}
\end{document}